\documentclass[english]{article}
\usepackage{graphicx}
\usepackage{epsfig}
\usepackage{authblk}
\usepackage{subfigure}
\usepackage{wrapfig}
\usepackage[T1]{fontenc}
\usepackage[latin9]{inputenc}
\usepackage{units}
\usepackage{amsmath}
\usepackage{amssymb}
\usepackage{esint}
\usepackage{babel}
\usepackage{color}
\usepackage{float}
\usepackage{lipsum}  
 \title{{\normalsize under consideration for publication in AIAA Journal} \vspace {7mm} \\
 Effect of Wall Transpiration and Heat Transfer on Nonlinear G\"{o}rtler Vortices in High-speed Boundary Layers}

\author[1]{Adrian Sescu}
\author[1]{Radwa Alaziz}
\author[2]{Mohammed Afsar}

\affil[1]{Department of Aerospace Engineering, Mississippi State University, MS 39762}
\affil[2]{Department of Mechanical \& Aerospace Engineering, Strathclyde University, 75 Montrose St. Glasgow, G1 1XJ, UK}

\begin{document}

\date{}

\maketitle

\begin{abstract}

G\"{o}rtler vortices in boundary layer flows over concave surfaces are caused by the imbalance between centrifugal effects and radial pressure gradients. Depending on various geometrical and/or freestream flow conditions, vortex breakdown via secondary instabilities leads to early transition to turbulence. It is desirable, therefore, to reduce vortex energy in an attempt to delay the transition from laminar to turbulent flow, and thereby achieve a reduced frictional drag. To this end, we apply a proportional control algorithm aimed at reducing the wall shear stress and the energy of G\"{o}rtler vortices evolving in high-speed boundary layers. The active control scheme is based on wall transpiration with sensors placed either in the flow or at the wall. In addition, we evaluate the effect of wall cooling and heating on G\"{o}rtler vortices evolving in high-speed boundary layers, by reducing or increasing the upstream wall temperature. The numerical results are obtained by solving the full Navier-Stokes equations in generalized curvilinear coordinates, using a high-order numerical algorithm. Our results show that the active control based on wall transpiration reduces both the wall shear stress and the energy of the G\"{o}rtler vortices; the passive control based on wall cooling or heating reduces the wall shear stress, but slightly increases the energy of the vortices in both supersonic and hypersonic regimes. 

\end{abstract}

\section{Introduction}

Streaks in pre-transitional boundary layer flows over flat or curved surfaces can form when the height of roughness elements exceeds a certain critical value (e.g., Choudhari \& Fischer \cite{choudhari}, White \cite{white1}, White et al. \cite{white2}, Goldstein et al. \cite{Goldstein1,Goldstein2,Goldstein3}, or Ruban \& Kravtsova \cite{Ruban}), or the amplitude of the free-stream disturbances is greater than a given threshold (e.g., Kendall \cite{Kendall}, Westin et al.  \cite{Westin}, Wu \& Choudhari \cite{wu1}, Matsubara \& Alfredsson \cite{Matsubara}, Leib et al. \cite{Leib}, Jacobs \& Durbin \cite{Jacobs}, Zaki \& Durbin \cite{Zaki}, Goldstein \& Sescu \cite{Goldstein4}, Ricco et al. \cite{ricco}, or Marensi et al. \cite{Marensi}). The streamwise velocity component exhibits elongated streaky features, characterized by adjacent regions of acceleration (high-speed regions) and deceleration (low-speed regions) of fluid particles (e.g., Kendall \cite{Kendall}, Matsubara \& Alfredsson \cite{Matsubara}, Landahl \cite{Landahl}, or Jacobs \& Durbin \cite{Jacobs}). Elongated streaks also develop inside a boundary layer flow along a concave surface caused by the imbalance between radial pressure gradients and centrifugal forces, and initiated by upstream disturbances (e.g., G\"{o}rtler \cite{Gortler}, Hall \cite{hall1,hall2,hall3}, Floryan and Saric \cite{Floryan}, Swearingen \& Blackwelder \cite{Swearingen}, Malik \& Hussaini \cite{Malik}, Saric \cite{Saric}, Li \& Malik \cite{Li}, Boiko et al. \cite{Boiko}, Wu et al. \cite{wu}, or Sescu et al. \cite{Sescu3}, Dempsey et al. \cite{Dempsey}, Xu et al. \cite{Xu}). For highly curved walls, for example, vortex formation occurs more rapidly and can significantly alter the mean flow causing the laminar flow to breakdown into turbulence. Under certain conditions, for example, G\"{o}rtler vortices can be efficient precursors to transition: they consist of counter-rotating streamwise vortices that grow at a certain rate, depending on the surface curvature and the receptivity of the boundary layer to environmental disturbances and surface imperfections. In compressible flows, G\"{o}rtler vortices have been studied in various contexts, as for example, El-Hady and Verma \cite{El-Hady}, Chen et al. \cite{Chen}, Hall and Fu \cite{hall4}, Fu and Hall \cite{Fu}, Dando and Seddougui \cite{Dando}, or Ren \& Fu \cite{Ren}.

It was recognized in many studies that boundary layer streaks are important in the path of transition to turbulence in a laminar flow; they are also important in the dynamics of turbulent boundary layers. Owing to their significance, it is desirable to reduce the streak energy, in an attempt to delay early nonlinear breakdown and transition. Previous studies and results indicated that the streaky structures in boundary layers are responsible for variations in the frictional drag or heat transfer, so any effective method of control must focus on restricting the development of these streaks. The next literature review in the following paragraphs focuses exclusively on boundary layer control based on wall blowing and suction. 

Wall blowing and suction - perhaps the most popular technique utilized to control boundary layers - can be applied via localized suction regions below low-velocity streaks and blowing regions below high-velocity streaks. Active wall control has been applied in the context of turbulent channel flow, targeting the reduction of the skin friction drag. Choi et al. \cite{Choi} conducted direct numerical simulations (DNS) of active wall control based on wall transpiration, with flow indicators located in a sectional plane parallel to the wall. As a result of this control technique, which was termed 'opposition control' by the authors, a frictional drag reduction of approximately 25\% was achieved. In the same study, Choi et al. \cite{Choi} investigated the same control algorithm except the sensors were placed at the wall, and the information was based on the leading term in the Taylor series expansion of the vertical component of velocity near the wall; a reduction of only 6\% was achieved. A somewhat similar control algorithm was applied by Koumoutsakos \cite{Koumoutsakos1,Koumoutsakos2}, with the control informed by the flow quantities from the wall. A more significant skin friction reduction (approximately 40\%) was obtained by using the vorticity flux components as inputs to the control algorithm. 

Lee et al. \cite{Lee} derived new suboptimal feedback control laws based on blowing and suction to manipulate the flow structures in the proximity to the wall, using information from the wall in the form of pressure or shear stress distribution. The reduction in the frictional drag was in the order of 16-20\%. Observing that the opposition control technique is more effective in low Reynolds number turbulent wall flows, Pamies et al. \cite{Pamies} proposed the utilization of the blowing only at high Reynolds numbers, and they obtained significant reduction in the skin-friction drag. Recently, Stroh et al. \cite{Stroh} conducted a comparison between the opposition control applied in the framework of turbulent channel flow and a spatially developing turbulent boundary layer, and found that the rates in the frictional drag reduction is approximately similar. An overview of the issues and limitations associated with the opposition control is given in the review article of Kim \cite{Kim}. 

Passive control mechanisms in the context of G\"{o}rtler vortices developing along curved walls have been considered and tested in a number of previous studies. Floryan and Saric \cite{Floryan2} studied the effect of suction on the G\"{o}rtler instabilities, and showed that the growth of these instabilities is reduced considerably, although the level of suction is much higher than what would be necessary to stabilize Tollmien-Schlighting waves. Floryan \cite{Floryan3} then extended this analysis to include 'a self-similar' blowing. The effect of wall suction and wall cooling on the growth and development of G\"{o}rtler instabilities in the compressible regime was investigated by El-Hady and Verma \cite{El-Hady2}. It was observed that at low suction or cooling rates the critical G\"{o}rtler number decreases, while at higher levels of suction or cooling there is an overall stabilizing effect, reducing the amplitude of the instabilities. Xu et al. \cite{Xu2} used direct numerical simulations to study the turbulent curved channel flow, and attempted to attenuate the G\"{o}rtler-Taylor vortices by means of a one-dimensional active cancellation approach. Both the attenuation of the growth of vortices and the reduction of the skin friction drag were achieved, with an overall drag reduction of approximately 12\%. 

In the compressible regime there have not been as many attempts to control the laminar or turbulent boundary layer as for incompressible flows. Some of these are mentioned next. Nishihara et al. \cite{Nishihara} performed experiments aimed at controlling a supersonic boundary layer, including flows of air and nitrogen, by using pulsed high-voltage discharge in the presence of magnetic fields. Im et al. \cite{Im} showed that a Mach 4.7 turbulent boundary layer can be manipulated by using a surface dielectric barrier discharge actuator. They observed that the boundary layer becomes thinner when the spanwise momentum is induced by a low-power actuator pair in the direction of the freestream flow. Xu et al. \cite{Xu2} employed wall suction in the framework of parabolized stability equations and Floquet thery to control the secondary instabilities in a boundary layer over a swept wing. They observed that as the suction velocity is increased the efficiency of the control algorithm increases, and that for a very large suction velocity the crossflow vortices are suppressed almost entirely. Sun et al. \cite{Sun} performed direct numerical simulations of compressible boundary layers, in both subsonic and supersonic regimes, under the effect of active dimples on the surface. A drag reduction of 15\% was achieved in the subsonic compressible flow, and 12\% reduction in the supersonic flow. Other studies focused on controlling the shock wave boundary layer interaction (see for example, Blinde et al. \cite{Blinde}, Pasquariello et al. \cite{Pasquariello}, Fukuda et al. \cite{Fukuda}, Vadillo et al. \cite{Vadillo}, Verma \& Hadjadj \cite{Verma}). The control techniques applied in the framework of hypersonic boundary layers is reviewed briefly by Fedorov \cite{Fedorov}.


The focus of this paper is on the manipulation of G\"{o}rtler vortices that develop in high-speed boundary layers, by using both active and passive control techniques based on wall transpiration and wall cooling, respectively. The former is performed using a proportional controller that updates the distributed wall blowing and suction, with the input being the streamwise velocity disturbance in a plane section that is parallel to the wall or the wall shear stress distribution. The wall transpiration is aimed at injecting or extracting momentum from the flow in the vertical direction, according to information that is provided from the wall shear stress or streamwise velocity component in a plane parallel to the wall. The use of this type of controller for flow control has been previously investigated in a number of studies (see, for example, Joshi et al. \cite{Joshi}, Hanson et al. \cite{Hanson} or Jacobson and Reynolds \cite{Jacobson}). Both supersonic and hypersonic boundary layers, with Mach number ranging from $1.5$ to $7.5$, are investigated using a high-order numerical algorithm that solves the full Navier-Stokes equations written in generalized curvilinear coordinates. Results show that while the control based on wall transpiration reduces both the wall shear stress and the energy of the G\"{o}rtler vortices, the wall cooling reduces the wall shear stress but increases the energy of the vortices in both supersonic and hypersonic boundary layers.

In section II, the mathematical model is introduced and described, including a discussion of the scalings of various independent or dependent variables, the appropriate initial and boundary conditions, as well as the numerical algorithm. Section III describes the control algorithm that is utilized to manipulate the energy of the G\"{o}rtler vortices and the wall shear stress. In section IV, results for various freestream Mach numbers, in both supersonic and hypersonic regimes, and for adiabatic and isothermal walls are included. Concluding remarks are given in section V.

\section{Problem formulation and numerical algorithm}\label{}

\subsection{Scalings}

Although this problem can be solved using a parabolized form of the Navier-Stokes equations with the curvature taken into account via a simple curvilinear transformation, here the governing equations consist of the full Navier-Stokes equations written in curvilinear coordinates. 
The vortex development up to the onset of secondary instabilities only is studied.

The equations considered here involve a generalized curvilinear coordinate transformation, which is written in the three-dimensional form as
$\xi = \xi \left(x,y,z \right),
\eta = \eta \left(x,y,z \right),
\zeta = \zeta \left(x,y,z \right)$,
where $\xi$, $\eta$ and $\zeta$ are the spatial coordinates in the computational space corresponding to the streamwise, wall-normal and spanwise directions, and $x$, $y$ and $z$ are the spatial coordinates in physical space; in this study, $\zeta$ is equivalent to $z$. This transformation allows for the mapping of the solution from the computational to the physical space and vice-versa. All dimensional spatial coordinates are normalized by the spanwise separation $\lambda^*$ of the G\"{o}rtler vortices, i.e.,

\begin{eqnarray}\label{NS}
(x,y,z) = \frac{(x^*,y^*,z^*)}{\lambda^*},
\end{eqnarray}
the velocity is scaled by the freestream velocity magnitude $V_{\infty}^*$, 

\begin{eqnarray}\label{NS}
(u,v,w) = \frac{(u^*,v^*,w^*)}{V_{\infty}^*},
\end{eqnarray}
the pressure by the dynamics pressure at infinity, $\rho_{\infty}^* V_{\infty}^{*2}$, and temperature by the freestream temperature, $T_{\infty}^*$. Reynolds number based on the spanwise separation, Mach number and Prandtl number are defined as

\begin{eqnarray}\label{NS}
R_{\lambda} = \frac{\rho_{\infty}^* V_{\infty}^* \lambda^*}{\mu_{\infty}^*}, \hspace{5mm}
M_{\infty} = \frac{V_{\infty}^*}{a_{\infty}^*}, \hspace{5mm}
Pr = \frac{\mu_{\infty}^* C_p}{k_{\infty}^*},
\end{eqnarray}
where $\mu_{\infty}^*$, $a_{\infty}^*$ and $k_{\infty}^*$ are freestream dynamic viscosity, speed of sound and thermal conductivity, respectively, and $C_p$ is the specific heat at constant pressure. All simulations are performed for air as an ideal gas, and no species equations are considered at high Mach numbers, such as the ones in the hypersonic regimes included in the analysis, $M=6$ and $M=7.5$, since it is expected that the effect of chemical reactions is negligible in the development of G\"{o}rtler vortices. The G\"{o}rtler number based on the boundary layer momentum thickness, $\theta^*$, is defined as

\begin{eqnarray}\label{NS}
G_{\theta} = R_{\theta} \sqrt{\frac{\theta^*}{r^*}},
\end{eqnarray}
where $R_{\theta}$ is the Reynolds number based on the momentum thickness and freestream velocity, and $r^*$ is the radius of the curvature.

\subsection{Governing equations}

 In conservative form, the Navier-Stokes equations are written as

\begin{eqnarray}\label{NS}
\mathbf{Q}_t
+ \mathbf{F} _{\xi}
+ \mathbf{G}_{\eta}
+ \mathbf{H}_{\zeta}
= \mathbf{S},
\end{eqnarray}
where subscript denote derivatives, the vector of conservative variables is given by

\begin{equation}
\mathbf{Q} = \frac{1}{J} \{ 
\begin{array}{rrrrrr}
\rho,   \hspace{4mm}
\rho u_i,   \hspace{4mm}
E
\end{array}
\}^{T}, i = 1,2,3,
\end{equation}
$\rho$ is the non-dimensional density of the fluid, $u_i = (u, v, w)$ is the non-dimensional velocity vector in physical space, and $E$ is the total energy. The flux vectors, $\mathbf{F}$, $\mathbf{G}$ and $\mathbf{H}$, are given by

\begin{eqnarray}
\mathbf{F} = \frac{1}{J} \left\{ 
\begin{array}{c}
\rho U,   \hspace{4mm}
\rho u_iU + \xi_{x_i} (p + \tau_{i1}),   \hspace{4mm}
E U + p U +  \xi_{x_i} \Theta_i
\end{array}
\right\}^{T}, \\
\mathbf{G} = \frac{1}{J} \left\{ 
\begin{array}{c}
\rho V,   \hspace{4mm}
\rho u_iV + \eta_{x_i} (p + \tau_{i2}),    \hspace{4mm}
E V + p V +  \eta_{x_i} \Theta_i
\end{array}
\right\}^{T},  \\ 
\mathbf{H} = \frac{1}{J} \left\{ 
\begin{array}{c}
\rho W,   \hspace{4mm}
\rho u_iW + \zeta_{x_i} (p + \tau_{i3}),    \hspace{4mm}
E W + p W +  \zeta_{x_i} \Theta_i
\end{array}
\right\}^{T},
\end{eqnarray}
where the contravariant velocity components are given by
$
U = \xi_{x_i} u_i ,   
V = \eta_{x_i} u_i,    
W = \zeta_{x_i} u _i,
$
with the Einstein summation convention applied over $i = 1,2,3$, the shear stress tensor and the heat flux are given as

\begin{equation}
\tau_{ij} = \frac{\mu}{Re} \left[
\left(
\frac{\partial \xi_k}{\partial x_j}  \frac{\partial u_i}{\partial \xi_k}  +
\frac{\partial \xi_k}{\partial x_i}  \frac{\partial u_j}{\partial \xi_k}
\right)
- \frac{2}{3} \delta_{ij} \frac{\partial \xi_l}{\partial x_k}  \frac{\partial u_k}{\partial \xi_l}
\right],
\end{equation}

\begin{equation}
\Theta_{i} = 
 u_j \tau_{ij} + \frac{\mu}{(\gamma-1)M_{\infty}^2 Re Pr}
\frac{\partial \xi_l}{\partial x_i}  \frac{\partial T}{\partial \xi_l},
\end{equation}
respectively, and $\mathbf{S}$ is the source vector term. 

The pressure $p$, the temperature $T$  and the density of the fluid are related through the equation of state, $p = \rho T / \gamma M_{\infty}^2$ under the hypothesis that the flow is non-chemically-reacting. The Jacobian of the curvilinear transformation from the physical space to computational space is denoted by $J$, and the derivatives $\xi_x$, $\xi_y$, $\xi_z$, $\eta_x$, $\eta_y$, $\eta_z$, $\zeta_x$, $\zeta_y$, and $\zeta_z$ represent grid metrics. The dynamic viscosity $\mu$ and thermal conductivity $k$ are linked to the temperature using the Sutherland's equations that are written in dimensionless form as

\begin{eqnarray}
\mu = T^{3/2} \frac{1 + C_1/T_{\infty}}{T+C_1/T_{\infty}}, \hspace{4mm}
k = T^{3/2} \frac{1 + C_2/T_{\infty}}{T+C_2/T_{\infty}},
\end{eqnarray}
where $C_1 = 110.4 K$, $C_2 = 194 K$ are constants, and $T_{\infty}$ is a reference temperature (Tannehill et al. \cite{Tannehill}).

\subsection{Numerical framework}

The numerical results are obtained by applying a high-order numerical algorithm to the compressible Navier-Stokes equations written in generalized curvilinear coordinates. Although the results reported in this study represent steady G\"{o}rtler vortices, the time dependent equations are solved to converge the solution to the steady state, with the time integration being performed via a third order TVD Runge-Kutta method~\cite{Liu}. The spatial derivatives in the wall-normal and spanwise directions are discretized using dispersion-relation-preserving schemes of Tam and Webb \cite{Tam}. Since the simulations included here are all in the supersonic regime, upwind schemes are utilized in the streamwise $\xi$ direction to avoid the imposition of an outflow condition since no waves are propagating back into the domain according to the characteristic velocities. 

\begin{figure}[H]
 \begin{center}
    \includegraphics[width=5cm]{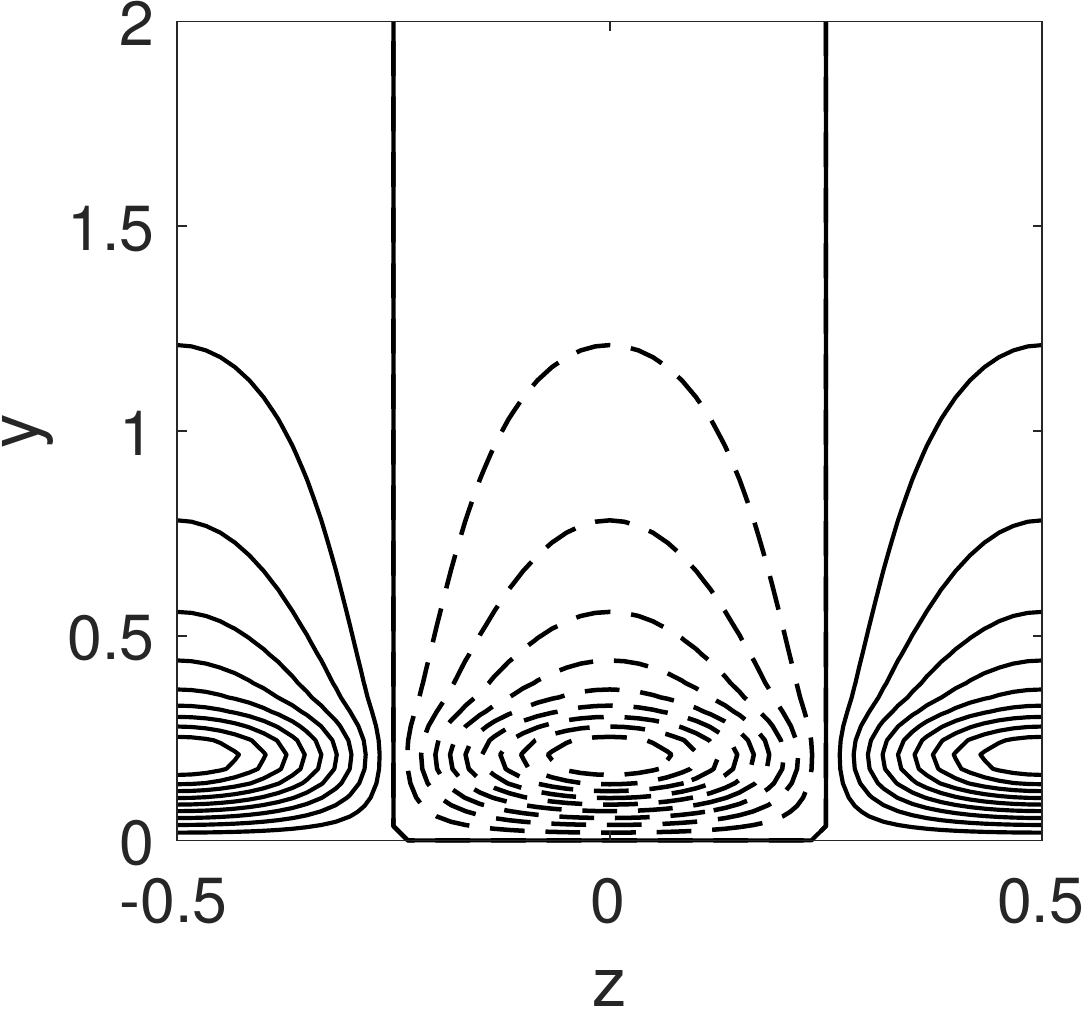}
    \includegraphics[width=5cm]{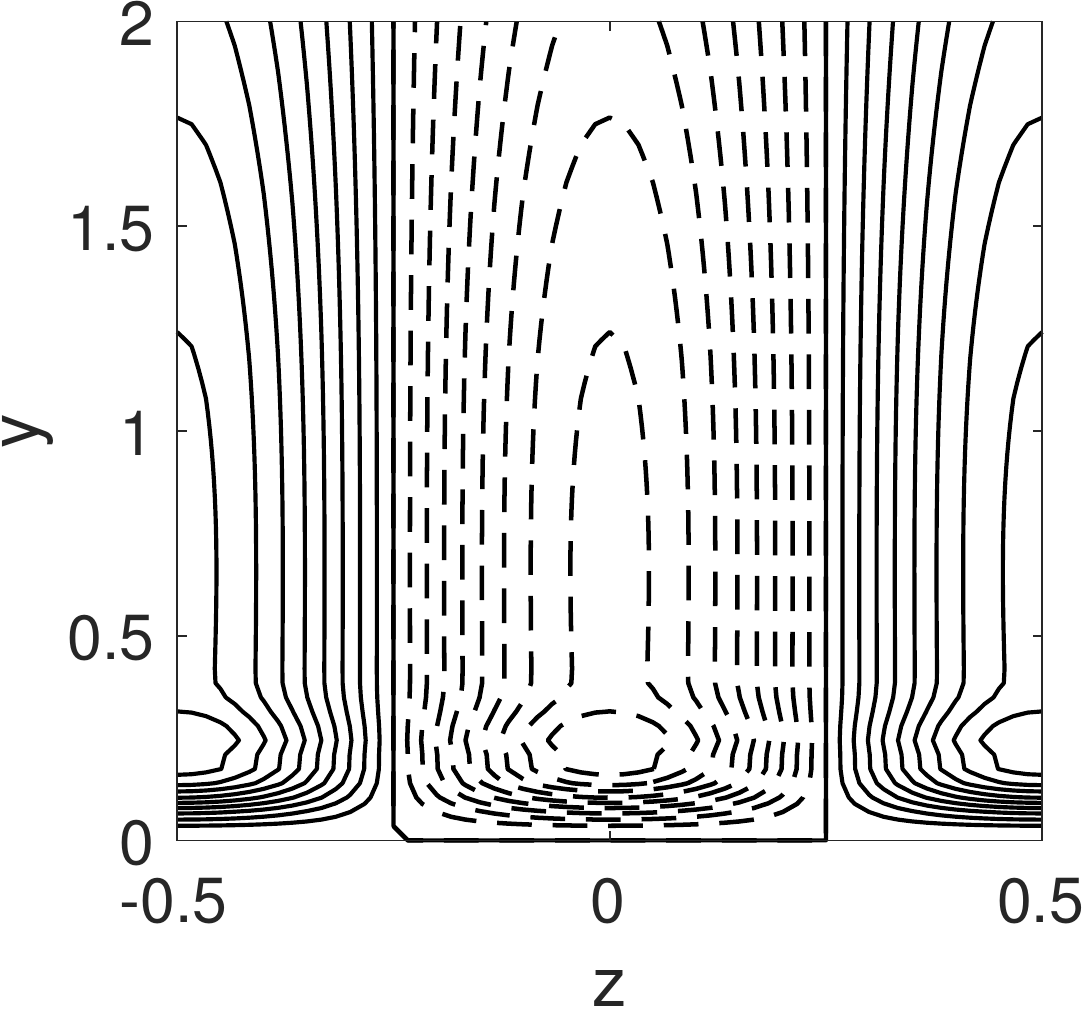}
    \includegraphics[width=5cm]{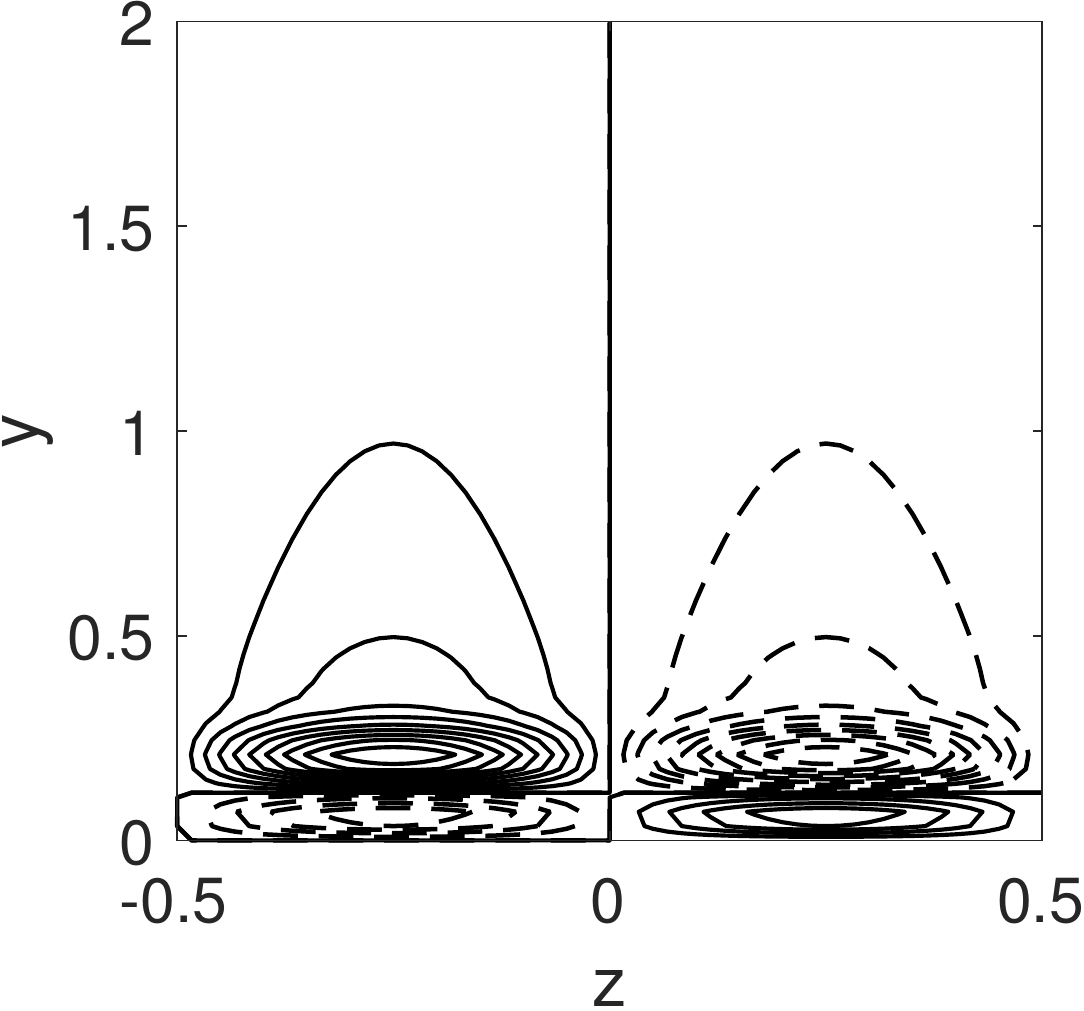}\\
a) \hspace{47mm} b) \hspace{47mm} c)
 \end{center}
  \caption{\label{} Contours of the inflow disturbance: a) $u'$ in the range $[-0.03,0.03]$; b) $v'$ in the range $[-0.002,0.002]$; c) $w'$ in the range $[-0.01,0.01]$.}
  \label{f1}
\end{figure}

The mean inflow condition is obtained separately from a precursor two-dimensional simulation, where a Blasius type boundary condition is imposed in the upstream. Wall-normal profiles of flow variables from a downstream location of the two-dimensional flow domain are imposed at the inflow of the three-dimensional domain, at each spanwise grid line. In order to excite the formation of G\"{o}rtler vortices in the downstream, a small velocity disturbance is superposed on the mean flow at the inflow boundary, with amplitude level equal to 0.3\% of the local base flow and functional form similar to the upstream disturbance imposed in the incompressible boundary layer analysis in Sescu \& Thompson \cite{Sescu3} (see also Goldstein et al. \cite{Goldstein1}). Contours of the inflow disturbance are shown in figure \ref{f1} for the three components of velocity, where the solid lines represent positive values and the dashed lines represent negative values. The vertical extent of the disturbance is scaled by the local boundary layer thickness for each Mach number considered in this study. The upstream disturbance should not affect the compressible solution considered in this paper since the amplitude of the disturbance is small, which means that it is unlikely to generate significant acoustic or entropy waves. Another effective approach to excite G\"{o}rtler instabilities is by imposing freestream disturbances as was done, for example, by Wu et al. \cite{wu}, Schrader et al. \cite{Schrader}, Marensi et al. \cite{Marensi}, or Xu et al. \cite{Xu2}; this will be the subject of a future study, wherein we plan to investigate both steady and unsteady G\"{o}rtler vortices. No slip boundary condition for velocity and adiabatic or isothermal conditions for temperature are imposed at the solid surface, and far field condition (vanishing gradients) are imposed at the top boundary. 

\section{Control Algorithm}

Both passive and active control algorithms are implemented and applied in the framework of evolving G\"{o}rtler vortices; the target is the reduction of the vortex energy or wall shear stress, in order to delay the transition to turbulence and/or decrease the frictional drag. The passive control scheme is based on wall cooling or heating that is applied downstream of the inflow boundary from some specified streamwise location. Up to this location the temperature of the wall is equal to the ambient temperature, which is set at $300$ K. Since the cooling or heating is imposed just downstream of the inflow boundary, the inflow condition is not affected by variations of temperature; this ensures that the G\"{o}rtler vortices are initiated at the inflow identically for all cases. A ramping function is used to impose the wall temperature, with a smooth the transition from the inflow wall temperature to the imposed temperature.

For the active control, a proportional control algorithm is utilized to determine the wall blowing and suction $v_w$ that will minimize the G\"{o}rtler vortex energy or the wall shear stress. The control variable here is either the streamwise velocity disturbance in a $\eta = const$ plane at a specified height above the wall surface, or the wall shear stress. Within the control scheme, the blowing and suction at the wall is varied according to the flow sensors that are placed either in the boundary layer or at the wall, while the former is not practically realizable owing to the difficulty in placing sensors in the flow without compromising the flow, while the latter is, since sensors placed at the wall obviously do not.
A typical proportional controller is considered here as 
$
A(X,z) = K_p*e(X,z),
$
where $K_p$ is the proportional gain, and 

\begin{eqnarray}
e(X,z) = U(X,y_{c},z) - U_m(X,y_{c})
\end{eqnarray} 
is the error signal obtained in terms of either sensors placed in the flow at a specified distance from the wall, and is defined by the difference between the streamwise velocity disturbance $U(\xi,\eta_c,\zeta)$ obtained from the Navier-Stokes equations and the spanwise averaged velocity $U_m(X,y_c)$ ($U$ is the streamwise contravariant velocity). For sensors placed at the wall, the wall shear stress represents the a control variable, and the error signal is given as

\begin{eqnarray}
e(X,z) = \tau_w(X,y_{c},z) - (\tau_w)_m(X,y_{c}),
\end{eqnarray}
where $\tau_w$ is the wall shear stress calculated from the Navier-Stokes solution, and $(\tau_w)_m$ is the spanwise averaged wall shear stress. The amplitude and the distribution of the wall transpiration can be updated at each iteration based on the control signal as 
$
v_w(X,z) = v_w(X,z) + A(X,z).
$
The proportional gain, $K_p$, can be determined, for example, by the frequency response method of Ziegler and Nichols \cite{Ziegler}. According to this method, the controller must be initiated with a small value of $K_p$. The proportional gain must be adjusted until a response is obtained that produces continuous oscillations (known as the ultimate gain). The desired proportional gain will then be half of the ultimate gain. The control algorithm utilized in this study is similar to the algorithm applied in Sescu et al. \cite{Sescu1} in the incompressible regime.

\section{Results and Discussion}\label{}

\subsection{Preliminaries}

We focus our attention on numerical simulations of the Navier-Stokes equations aimed to scrutinize the effectiveness of the control schemes on G\"{o}rtler vortices developing in both supersonic and hypersonic boundary layers, where Mach number is varied between $1.5$ and $7.5$. Since there are no obstructions at the wall or significant streamwise variations of the mean flow no shock waves or other discontinuities are expected. It is worth mentioning that supersonic boundary layer flows over concave walls, free of discontinuities, have been previously considered in several studies such as, for example, Li et al. \cite{Li2} or Ren \& Fu \cite{Ren}. Note that the shock that would be posed by the leading edge is not taken into account in these simulations since the inflow boundary is placed downstream of the leading edge, so the effect from the shock is expected to be negligible. This justifies the reason for imposing the mean inflow condition via a precursor two-dimensional simulation as was described previously. Table \ref{t1} gives the flow parameters for the five cases considered in the analysis. We follow the trends that are common in the majority of studies pertaining G\"{o}rtler vortices, and choose the spanwise separation of the vortices to be in the same order of magnitude as the boundary layer thickness, such that the ratio between the two is set approximately the same among all Mach numbers. The G\"{o}rtler number range given in table \ref{t1} (i.e., between $20$ and $35$) commonly represents nonlinear G\"{o}rtler vortices.

\begin{table}[htpb]
 \begin{center}
  \caption{{\color{black} Flow parameters.}}
  \label{t1}
  \begin{tabular}{rrrrrr} \hline
Mach number, $M$ & Unit Reynolds number & G\"{o}rtler number, $G_{\theta}$ & {\color{black} Spanwise separation, $\Lambda^*$ [m]}  \\\hline
       {\color{black} $M = 1.5$}  &  {\color{black} 1.40E+6}  & {\color{black} 22.00}  & {\color{black} 0.0120}   \\
       {\color{black} $M = 3.0$}  &  {\color{black} 2.81E+6}  & {\color{black} 25.21}  & {\color{black} 0.0057}   \\
       {\color{black} $M = 4.5$}  &  {\color{black} 4.18E+6}  & {\color{black} 28.28}  & {\color{black} 0.0043}   \\
       {\color{black} $M = 6.0$}  &  {\color{black} 5.55E+6}  & {\color{black} 31.45}  & {\color{black} 0.0036}   \\
       {\color{black} $M = 7.5$}  &  {\color{black} 7.09E+6}  & {\color{black} 34.57}  & {\color{black} 0.0031}   \\
\hline
  \end{tabular}
 \end{center}
\end{table} 

{\color{black} As mentioned previously, the excitation of Gortler vortices is realized by a small disturbance that is superposed on the base mean flow at the inflow boundary.} Either the streamwise velocity disturbance $U'$ in a surface that is parallel to the wall or the wall shear stress, $\tau_w$, is considered as the input sensor for the control algorithm. The control is performed with wall transpiration and wall cooling {\color{black} or heating}, and the results in terms of streamwise velocity contours, profiles of streamwise velocity, as well as axial distributions of vortex energy and wall shear stress are compared to results from a boundary layer without control.

\begin{figure}[H]
 \begin{center}
    \includegraphics[width=13cm]{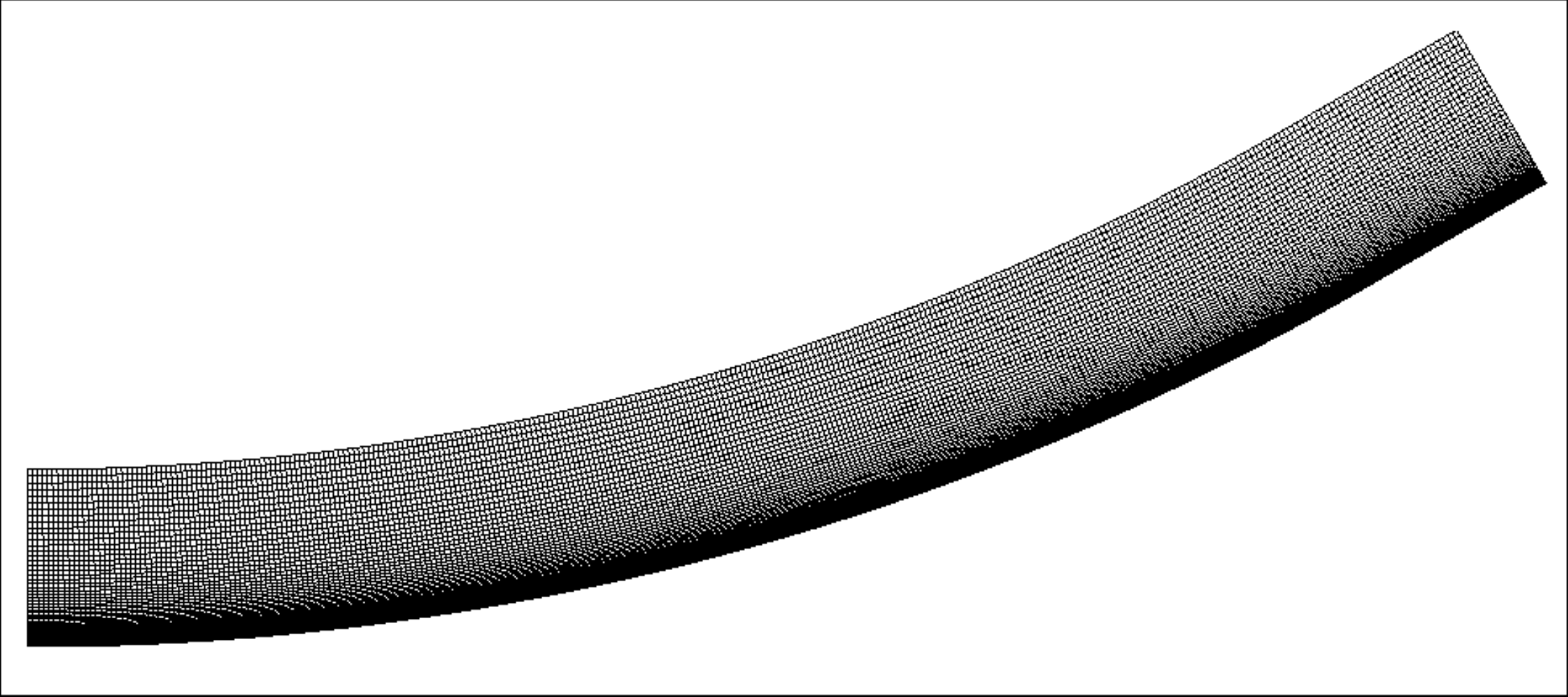}
 \end{center}
  \caption{\label{} Mesh configuration.}
  \label{f2}
\end{figure}

\begin{figure}[H]
 \begin{center}
    \includegraphics[width=7cm]{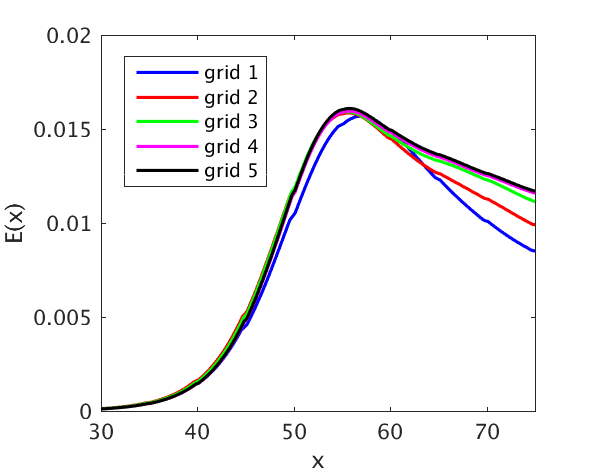}
    \includegraphics[width=7cm]{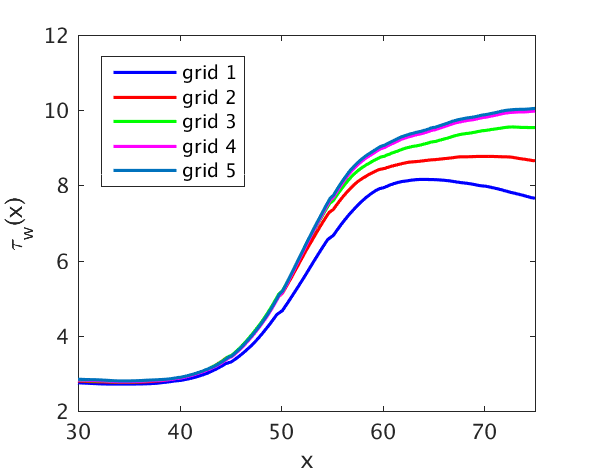} 
 \end{center}
  \caption{\label{} Vortex energy (left) and wall shear stress (right) for four grids of different resolutions.}
  \label{f3}
\end{figure}

{\color{black} To have confidence about our results, we perform} a grid convergence study to ensure that the results are not sensitive to the grid resolution. {\color{black} The side view of} the grid in figure \ref{f2}, {\color{black} shows} stretching that was applied in the radial direction towards the farfield boundary, {\color{black} with clustering of the grid points in proximity of the wall. The stretching in the radial direction ensures that the waves moving away from the wall are dissipated and do not return into the flow domain as spurious reflected waves. The mesh features} equally spaced grid points in the streamwise and spanwise directions. {\color{black} Within the grid study, five} meshes of different resolutions have been chosen {\color{black} - denoted $grid1$, $grid2$, $grid3$, $grid4$, and $grid5$ - and simulations corresponding to each Mach number have been performed as part of the grid convergence study. A systematic analysis of the grid resolution was performed to determine what direction in the computational space ($\xi$, $\eta$, or $\zeta$) has the most significant effect on the sensitivity of the results; it was found that first the streamwise $\xi$ and then the wall-normal $\eta$ directions had significant effects. Table \ref{t2} lists the total number of grid points, for all Mach numbers, that was utilized in the convergence analysis}. In the simulations, we have chosen the spanwise length of the domain to correspond to a single G\"{o}rtler vortex although subsequent contour plots in the next subsections will include two vortices in order to highlight the distance between two adjacent mushroom shapes. In figure \ref{f3}, distributions of the vortex energy and spanwise averaged wall shear stress as a function of the streamwise coordinate, {\color{black} for Mach number $4.5$}, show that the results obtained from $grid4$ and $grid5$ are very close to each other {\color{black} (similar conclusions hold for the other Mach numbers)}. As a result, the grid resolutions associated with $grid4$ were chosen for the {\color{black} numerical simulations that are discussed next.}

\begin{table}[htpb]
 \begin{center}
  \caption{{\color{black} Number of points in different grids.}}
  \label{t2}
  \begin{tabular}{rrrrrr} \hline
       {\color{black} Mach number} & {\color{black} $grid1$} & {\color{black} $grid2$} & {\color{black} $grid3$} & {\color{black} $grid4$} & {\color{black} $grid5$}  \\\hline
       {\color{black} $M = 1.5$}  &  {\color{black} 329,232}  & {\color{black} 990,000}  & {\color{black} 1,771,200}  & {\color{black} 2,763,600}  & {\color{black} 3,830,400} \\
       {\color{black} $M = 3.0$}  &  {\color{black} 329,232}  & {\color{black} 930,000}  & {\color{black} 1,684,800}  & {\color{black} 2,646,000}  & {\color{black} 3,696,000} \\
       {\color{black} $M = 4.5$}  &  {\color{black} 329,232}  & {\color{black} 870,000}  & {\color{black} 1,598,400}  & {\color{black} 2,528,400}  & {\color{black} 3,561,600} \\
       {\color{black} $M = 6.0$}  &  {\color{black} 329,232}  & {\color{black} 810,000}  & {\color{black} 1,512,000}  & {\color{black} 2,410,800}  & {\color{black} 3,427,200} \\
       {\color{black} $M = 7.5$}  &  {\color{black} 329,232}  & {\color{black} 750,000}  & {\color{black} 1,425,600}  & {\color{black} 2,293,200}  & {\color{black} 3,292,800} \\
\hline
  \end{tabular}
 \end{center}
\end{table} 

\subsection{Effect of wall transpiration}

In this section, we analyze the effect of distributed wall {\color{black} blowing and suction} on the development of G\"{o}rtler vortices, {\color{black} quantified} in terms of either the vortex energy or spanwise averaged wall shear stress {\color{black} in both the supersonic and hypersonic flow regimes, with flow conditions outlined in table \ref{t1}}. The amplitude of the incoming disturbance that excites the G\"{o}rtler vortex is $0.3\%$ of the mean flow{\color{black}, which is sufficiently large to excite nonlinear G\"{o}rtler vortices, and - at the same time - small enough to avoid the generation of significant spurious waves at the inflow boundary}. Both isothermal and adiabatic wall conditions are studied, where the wall temperature associated with the former is set equal to the ambient temperature, $300$ K. The length of the domain is set based on the {\color{black} streamwise distribution of the} vortex energy, with the outflow boundary {\color{black} positioned} well downstream of the energy saturation starting point. The sensors of the control scheme are placed either in the flow, where the streamwise velocity disturbance is {\color{black} monitored} in a surface located at {\color{black} $0.1\Lambda^*$} height from the wall, or at the wall, where the wall shear stress is monitored. {\color{black} In all numerical simulations a unique proportional gain $K_p=0.01$ was utilized for all five Mach numbers. The increase in the cost associated with applying the control is insignificant since the iterations within the control algorithm are the same as the iterations used to converge the solution to the steady state.} Results are discussed in terms of contour plots of the density in cross-sections through the vortex, energy distribution along the streamwise direction, spanwise averaged wall shear stress (which is an indication of the skin friction reduction), and profiles of the streamwise velocity in the center of the mushroom shape or in between. The wall transpiration is applied starting at the streamwise location $x=10$ (spanwise separation units) from the inflow boundary, {\color{black} by applying a ramping function in order to avoid the generation of discontinuities in the flow}.

\begin{figure}[H]
 \begin{center}
      \includegraphics[width=3.18cm]{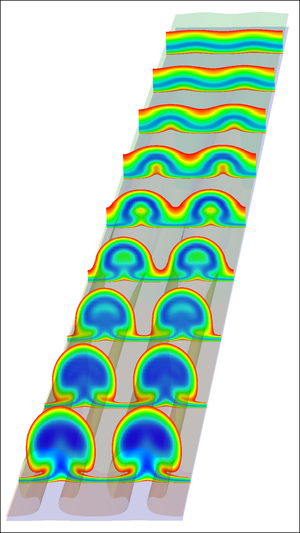}
      \includegraphics[width=3.18cm]{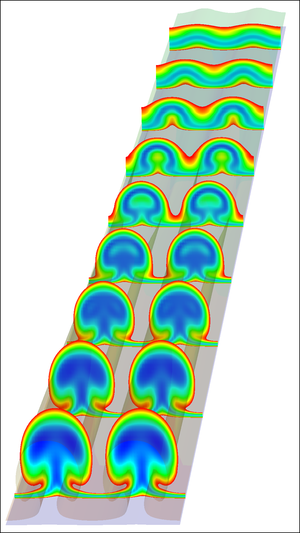}
      \includegraphics[width=3.18cm]{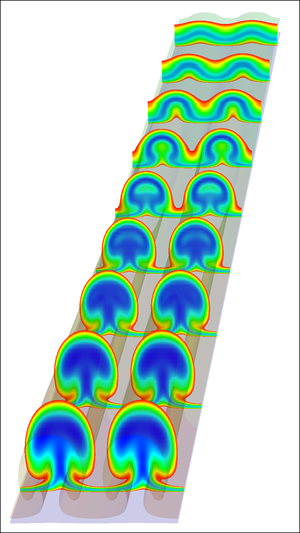}
      \includegraphics[width=3.18cm]{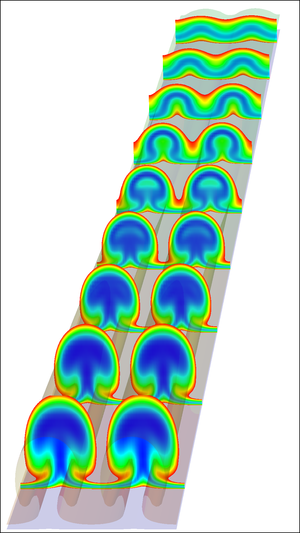}
      \includegraphics[width=3.18cm]{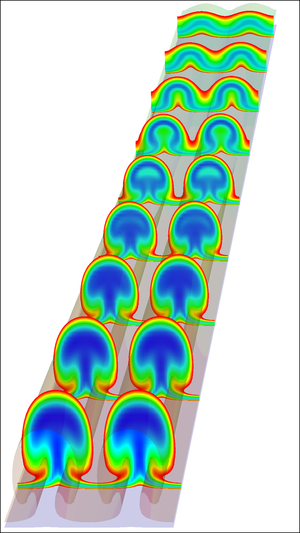}\\
$a_1$) \hspace{28mm} $b_1$)  \hspace{28mm} $c_1$)  \hspace{28mm} $d_1$)  \hspace{28mm} $e_1$) \\
      \includegraphics[width=3.18cm]{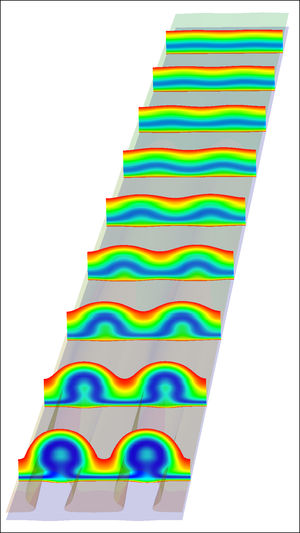}
      \includegraphics[width=3.18cm]{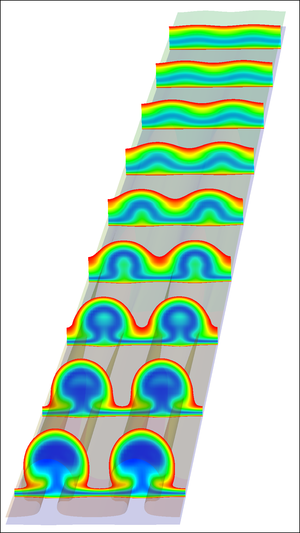}
      \includegraphics[width=3.18cm]{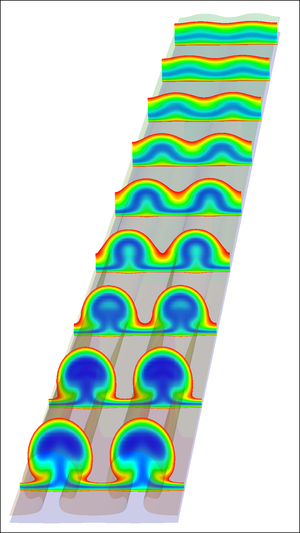}
      \includegraphics[width=3.18cm]{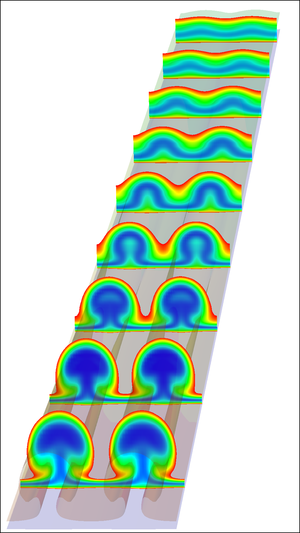}
      \includegraphics[width=3.18cm]{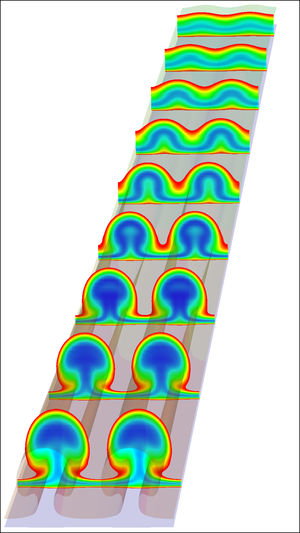}\\
$a_2$) \hspace{28mm} $b_2$)  \hspace{28mm} $c_2$)  \hspace{28mm} $d_2$)  \hspace{28mm} $e_2$)
      \includegraphics[width=3.18cm]{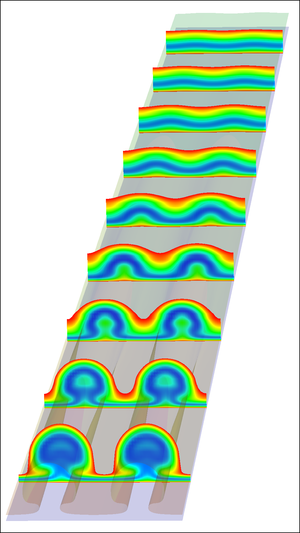}
      \includegraphics[width=3.18cm]{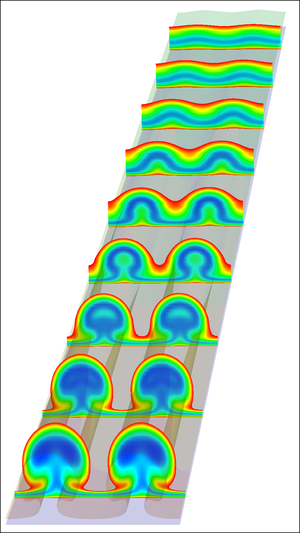}
      \includegraphics[width=3.18cm]{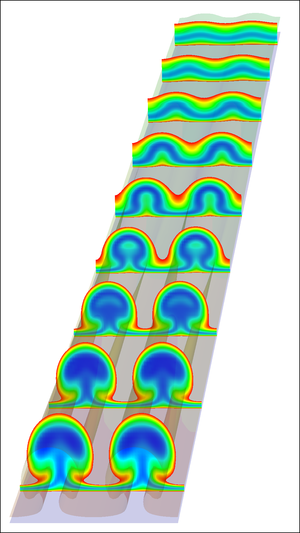}
      \includegraphics[width=3.18cm]{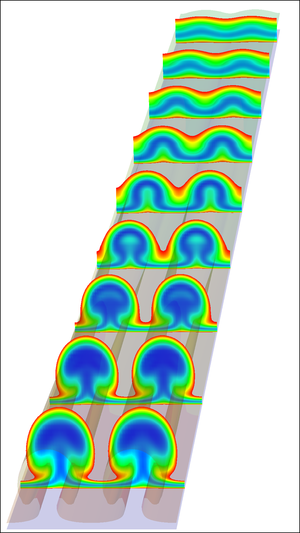}
      \includegraphics[width=3.18cm]{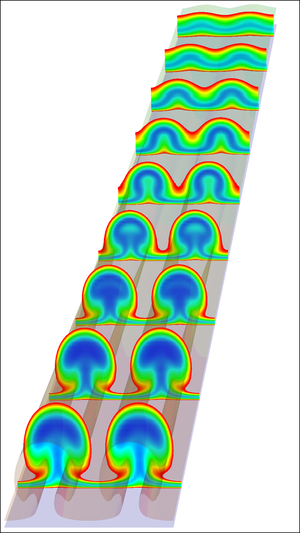}\\
$a_3$) \hspace{28mm} $b_3$)  \hspace{28mm} $c_3$)  \hspace{28mm} $d_3$)  \hspace{28mm} $e_3$)
 \end{center}
  \caption{\label{} Density contour plots at different streamwise locations for the isothermal wall: a) $M=1.5$; b) $M=3.0$; c) $M=4.5$; d) $M=6.0$; e) $M=7.5$. Subscript 1: no control; Subscript 2: control based on $u$; Subscript 3: control based on $\tau_w$.}
  \label{f4}
\end{figure}

In figure \ref{f4}, we plot density contours in cross sections $(y,z)$ planes at different streamwise locations, starting from $x=10$ to $x=60$, corresponding to the adiabatic wall condition. The top row, which represents the uncontrolled flow with Mach number varied from $1.5$ (left) to $7.5$ (right), indicate that as the Mach number is increased the {\color{black} relative} size of the mushroom shape increases in the wall-normal directions. {\color{black} The word 'relative' is utilized here because the spanwise separations of vortices are different for each Mach number, so the vortices associated with high Mach numbers are actually smaller than the vortices associated with small Mach numbers when we consider physical units - see table \ref{t1}}. The spanwise separations were chosen sufficiently large {\color{black} with respect to the boundary layer thickness} to allow the vortices to fully develop even for the largest Mach number of $7.5$, unlike the study of Ren and Fu \cite{Ren} where the small spanwise separation hampered the formation of mushroom shapes for their largest Mach number of $6$. {\color{black} On the other hand, if the spanwise separation is much greater than the boundary layer thickness the wavenumber of the disturbances in the downstream may increase, that is secondary vortices may form between two main G\"{o}rtler vortices. } 

The second row of figure \ref{f4} corresponds to density contour for the controlled boundary layer based on sensors {\color{black} placed in the flow,} at a certain distance from the wall. The sensor in this case is the streamwise velocity disturbance - which {\color{black} can be regarded as a measure of} the streak {\color{black} amplitude} - {\color{black} at $y=0.1\Lambda^*$ from the wall}. The streak {\color{black} amplitude} is defined as the half the difference between the highest and the lowest streamwise velocities {\color{black} at two spanwise locations (Paredes et al. \cite{Paredes})}. Another option is to use the wall-normal velocity disturbance as in the previous studies that were focused on controlling turbulent boundary layers (e.g., the opposition control of Choi et al. \cite{Choi}), but it was found based on numerical experiments that the control based on the streamwise velocity is {\color{black} more effective in reducing the energy of} G\"{o}rtler vortices. {\color{black} One obvious observation inferred from inspecting the second row of figure \ref{f4} is the level of reduction of the mushroom size as a result of imposing the distributed blowing and suction. The reduction of the vortex strength seems to be more effective at small Mach numbers (below $3$), while at higher Mach numbers the reduction if slight smaller and the disturbances seem to retain the mushroom shape.} The third row corresponds to the control based on sensors placed on the wall, {\color{black} represented by the wall shear stress  distribution; this kind of control is practically realizable since the sensors do not interfere with the boundary layer flow}. {\color{black} However, for this type of sensors,} the reduction of the mushroom size is slightly {\color{black} less significant} when compared to the control based on the streamwise velocity disturbance (see the second row in figure \ref{f4}). {\color{black} This will also be revealed by vortex energy distributions in the subsequent figures.}

\begin{figure}[H]
 \begin{center}
      \includegraphics[width=3.18cm]{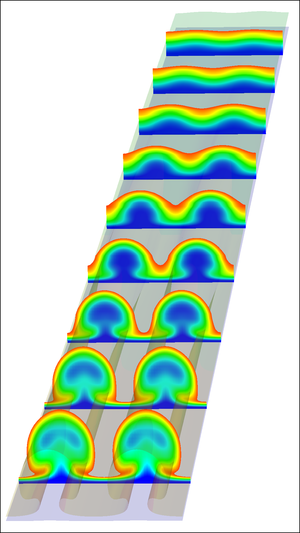}
      \includegraphics[width=3.18cm]{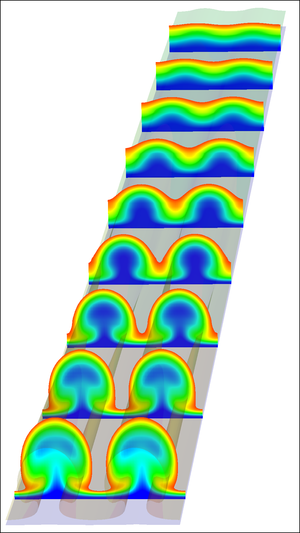}
      \includegraphics[width=3.18cm]{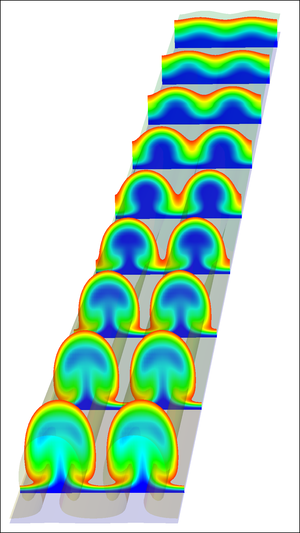}
      \includegraphics[width=3.18cm]{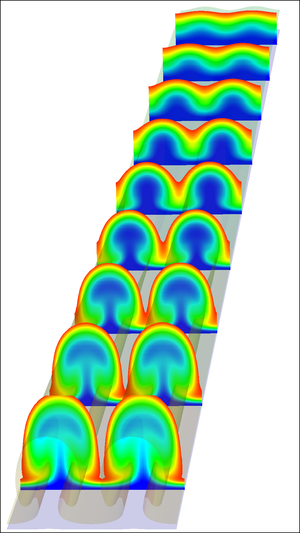}
      \includegraphics[width=3.18cm]{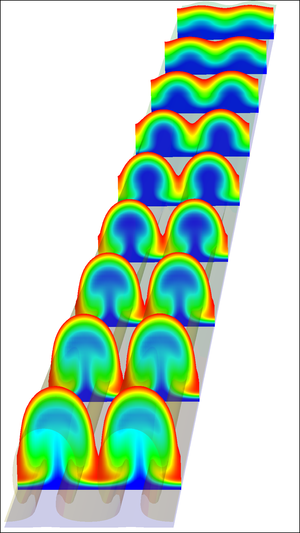}\\
$a_1$) \hspace{28mm} $b_1$)  \hspace{28mm} $c_1$)  \hspace{28mm} $d_1$)  \hspace{28mm} $e_1$) \\
      \includegraphics[width=3.18cm]{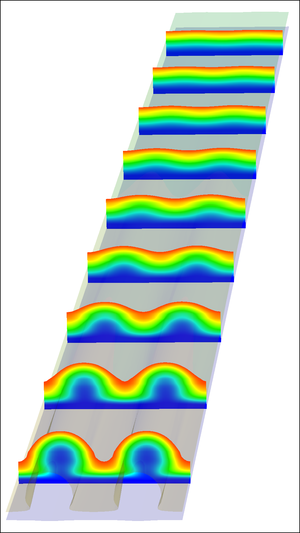}
      \includegraphics[width=3.18cm]{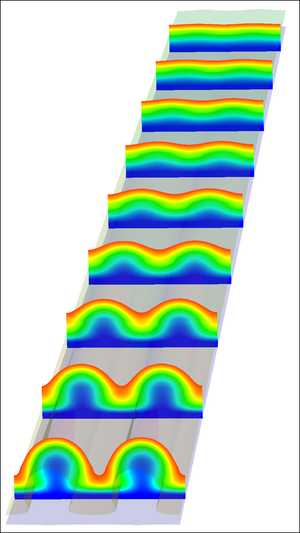}
      \includegraphics[width=3.18cm]{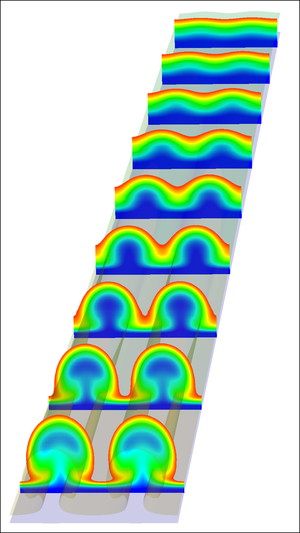}
      \includegraphics[width=3.18cm]{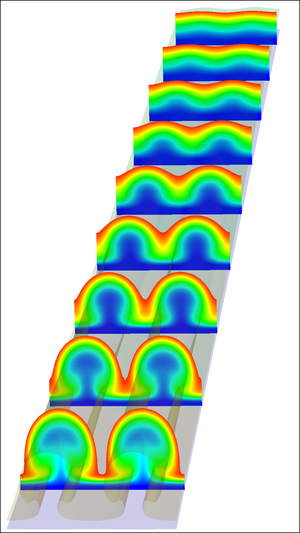}
      \includegraphics[width=3.18cm]{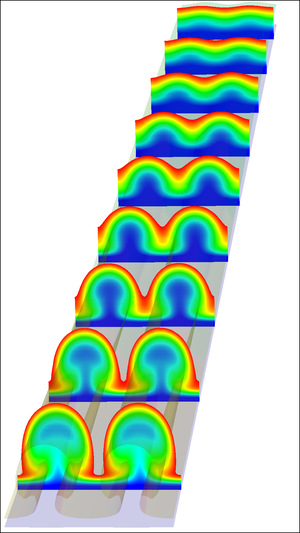}\\
$a_2$) \hspace{28mm} $b_2$)  \hspace{28mm} $c_2$)  \hspace{28mm} $d_2$)  \hspace{28mm} $e_2$)
      \includegraphics[width=3.18cm]{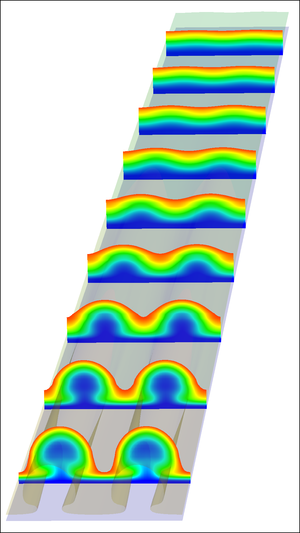}
      \includegraphics[width=3.18cm]{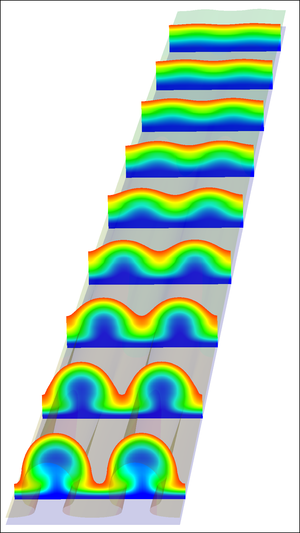}
      \includegraphics[width=3.18cm]{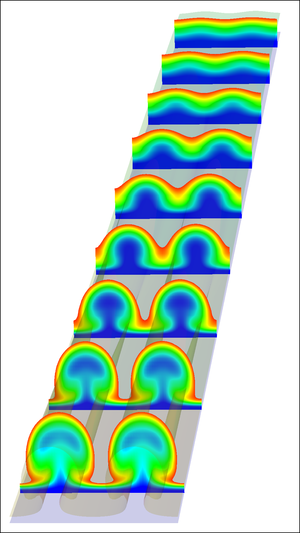}
      \includegraphics[width=3.18cm]{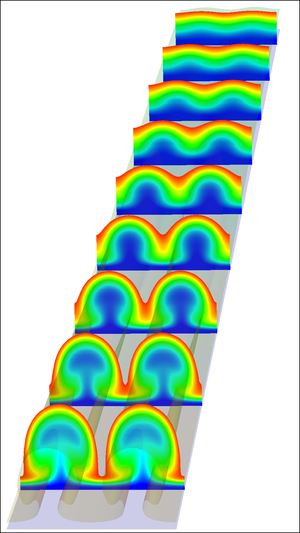}
      \includegraphics[width=3.18cm]{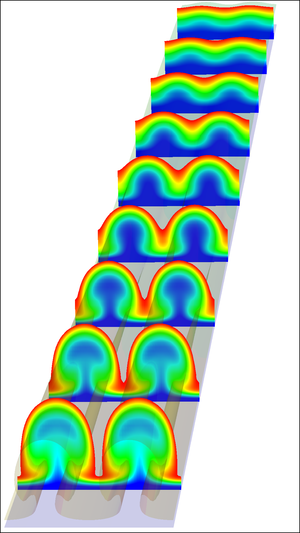}\\
$a_3$) \hspace{28mm} $b_3$)  \hspace{28mm} $c_3$)  \hspace{28mm} $d_3$)  \hspace{28mm} $e_3$)
 \end{center}
  \caption{\label{} Density contour plots at different streamwise locations for the adiabatic wall: a) $M=1.5$; b) $M=3.0$; c) $M=4.5$; d) $M=6.0$; e) $M=7.5$. Subscript 1: no control; Subscript 2: control based on $u$; Subscript 3: control based on $\tau_w$.}
  \label{f5}
\end{figure}

{\color{black} We plot similar sets} of density contour plots in figure \ref{f5} except {\color{black} the wall comes into play as an} adiabatic surface, {\color{black} which can trigger much higher temperatures inside the boundary layer, especially for high Mach numbers. As a result of the increased temperature,} the boundary layer thickness becomes larger {\color{black} than the corresponding thickness in the isothermal condition} since a temperature increase {\color{black} close to the wall implies} an increase in the {\color{black} local} effective viscosity, which adds more dissipation into the flow. {\color{black} As far as the control of vortices is concerned, however,} the level of vortex strength reduction as noticed in the second and third rows of figure \ref{f5}, corresponding to the control based on the streamwise velocity disturbance and wall shear stress, respectively, seems to be {\color{black} approximately} the same as is for the isothermal wall. {\color{black} This will be revealed more clearly in the energy plots that are presented next.}

\begin{figure}[H]
 \begin{center}
      \includegraphics[width=5.4cm]{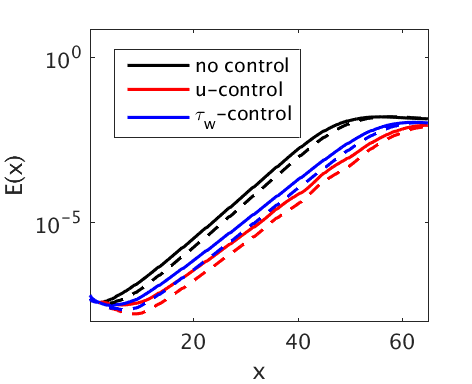} 
      \includegraphics[width=5.4cm]{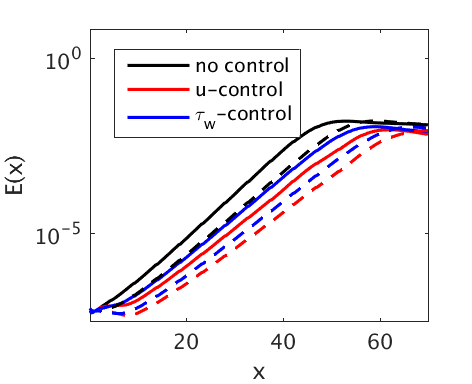}
      \includegraphics[width=5.4cm]{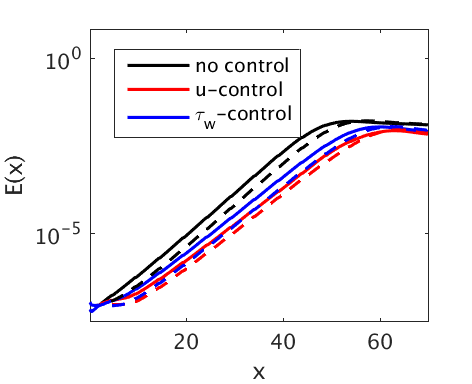}\\
a) \hspace{50mm} b) \hspace{50mm} c) \\
      \includegraphics[width=5.4cm]{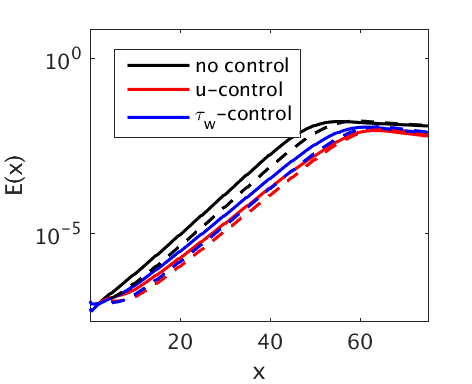}
      \includegraphics[width=5.4cm]{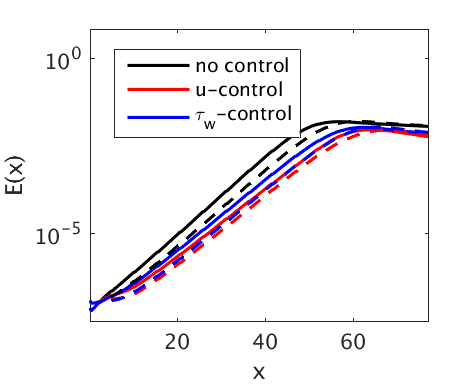}\\
d) \hspace{60mm} e) 
 \end{center}
  \caption{\label{} Energy of the disturbance for isothermal (solid line) and adiabatic (dashed line) wall: a) $M_{\infty} = 1.5$ b) $M_{\infty} = 3.0$; c) $M_{\infty} = 4.5$; d) $M_{\infty} = 6.0$; e) $M_{\infty} = 7.5$.}
  \label{f6}
\end{figure}

Here, we quantify the effect of the control schemes via {\color{black} streamwise} distributions of the vortex energy and spanwise averaged wall shear stress. {\color{black} The former gives indication to when the vortex will lead to secondary instability, while the latter provides some evidence of potential skin frictional drag reduction.} Figure \ref{f6} shows the kinetic energy {\color{black} evolution} of the disturbance, calculated {\color{black} as an integral along both wall-normal and spanwise directions} according to

\begin{equation}\label{jj}
E(x) = \intop_{z_1}^{z_2}  \intop_{0}^{\infty}  \left[ \left| u(x,y,z) - u_m(x,y) \right|^{2} +  \left| v(x,y,z) - v_m(x,y) \right|^{2} +  \left| w(x,y,z) - w_m(x,y) \right|^{2} \right] dzdy,
\end{equation}
where $u_m(x,y)$, $v_m(x,y)$, and $w_m(x,y)$ are the spanwise mean components of the velocity, and $z_1$ and $z_2$ are the coordinates of the boundaries in the spanwise {\color{black} periodic} direction. Control strategies with streamwise velocity {\color{black} disturbance sensors the flow} (red curves {\color{black} in figure \ref{f6}}) appear to be more effective in reducing the energy of the vortices: more than one order of magnitude reduction is noticed from these plots. Although all five cases in figure \ref{f6} show the same level of energy reduction, closer look indicates that the level of reduction decreases slightly as the Mach number is increased. {\color{black} Interestingly, however, the energy levels in the saturation region are in the same order of magnitude for both the uncontrolled and controlled cases. Comparing the solid lines to the dashed lines in figure \ref{f6} indicates that the wall cooling slightly increases the vortex energy compared to the adiabatic wall condition case.} The energy plots also reveal that the control scheme is slightly more effective in the adiabatic wall case for both types of sensors.

\begin{figure}[H]
 \begin{center}
      \includegraphics[width=5.4cm]{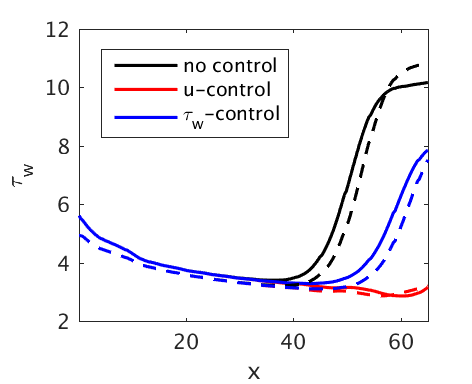}
      \includegraphics[width=5.4cm]{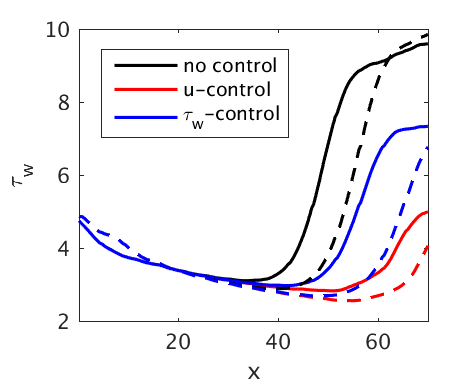}
      \includegraphics[width=5.4cm]{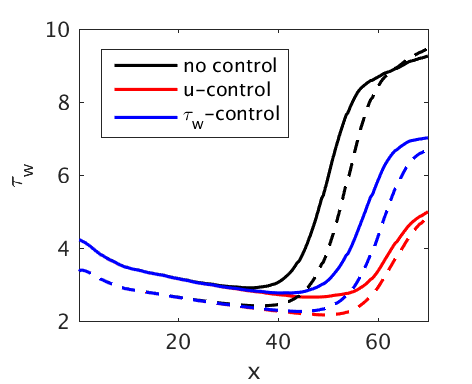}\\
a) \hspace{50mm} b) \hspace{50mm} c) \\
      \includegraphics[width=5.4cm]{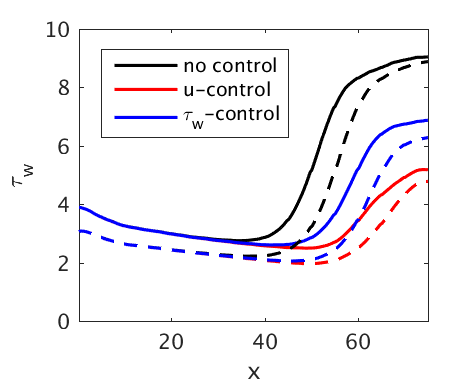}
      \includegraphics[width=5.4cm]{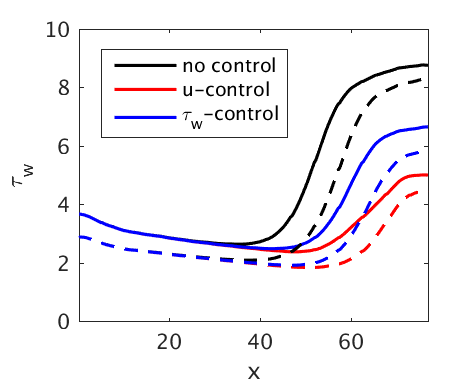}\\
d) \hspace{60mm} e) 
 \end{center}
  \caption{\label{} Spanwise averaged wall shear stress for isothermal (solid line) and adiabatic (dashed line) wall: a) $M_{\infty} = 1.5$ b) $M_{\infty} = 3.0$; c) $M_{\infty} = 4.5$; d) $M_{\infty} = 6.0$; e) $M_{\infty} = 7.5$.}
  \label{f7}
\end{figure}

In figure \ref{f7}, the spanwise averaged wall shear stress, which is calculated {\color{black} by integrating the velocity gradient in the spanwise direction as}

\begin{equation}\label{}
\tau_w(x) = \frac{1}{z_2-z_1} \intop_{z_1}^{z_2} \frac{\partial u_t}{\partial \eta} (x,0,z) dz,
\end{equation}
is represented as a function of the streamwise coordinate, where the black curves correspond to the uncontrolled case. Consistent with the distributions of energy in figure \ref{f6}, these show that the wall shear stress is reduced for both {\color{black} types of sensors}, with the control based on the streamwise velocity being more effective. The difference between the isothermal and the adiabatic wall conditions is also similar to the trends observed in the energy plots, but at low Mach number only. The wall shear stress for adiabatic conditions is lower compared to isothermal, upstream of the energy saturation point; however, the trend switches beyond the saturation point. For high Mach number cases, the wall shear stress for the adiabatic condition is lower than the wall shear stress for the isothermal wall at all streamwise locations, {\color{black} a result that was obtained in previous studies (Spall and Malik \cite{Spall} or Elliot \cite{Elliot})}. {\color{black} Overall, all vortex energy and wall shear stress plots suggest that the distributed wall blowing and suction is capable of reducing the strength of the vortices, thus delaying the secondary instability onset, and also reducing the wall shear stress, thus decreasing the frictional drag. This is valid in both supersonic and hypersonic regimes.}


\begin{figure}[H]
 \begin{center}
      \includegraphics[width=6.1cm]{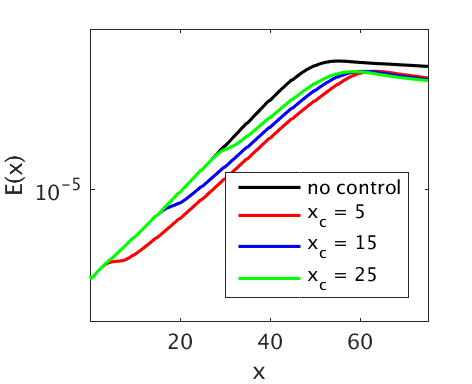}
      \includegraphics[width=6.1cm]{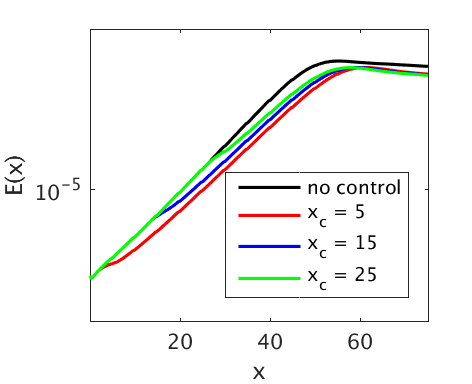} \\
a) \hspace{64mm} b) \\
 \end{center}
  \caption{\label{} {\color{black} Energy of the disturbance for different values of $x_c$ (Mach number is $6$): a) control based on $u$ b) control based on $\tau_w$.}}
  \label{f8}
\end{figure}

\begin{figure}[H]
 \begin{center}
      \includegraphics[width=6.1cm]{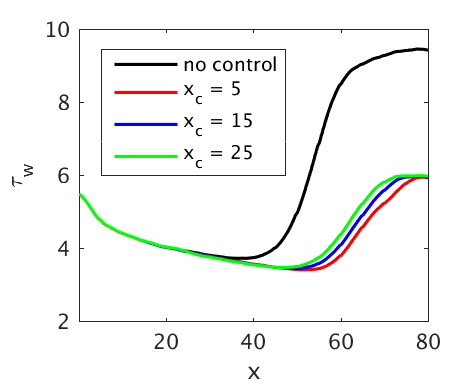}
      \includegraphics[width=6.1cm]{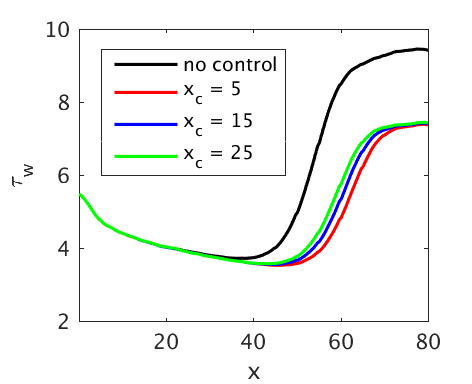} \\
a) \hspace{64mm} b) \\
 \end{center}
  \caption{\label{} {\color{black} Spanwise averaged wall shear stress for different values of $x_c$ (Mach number is $6$): a) control based on $u$ b) control based on $\tau_w$.}}
  \label{f9}
\end{figure}

{\color{black} We looked into the effect of changing the streamwise location $x_c$, where the control based on wall blowing and suction is initiated. Three different streamwise locations are considered: $x_c = 5, 15$ and $25$. Curves of vortex energy as a function of the streamwise coordinate plotted in figure \ref{f8} point out that the smaller the location $x_c$, the higher the energy reduction is, for both types of sensors. While the same conclusion is inferred with regard to the wall shear stress distribution plotted in figure \ref{f9}, the level of reduction in this case is not so significant (note that the energy plots in figure \ref{f8} are in logarithmic scale with respect to the vertical direction). In conclusion, the streamwise location $x_c$ may have a more significant effect on the secondary instability of the vortices, and less effect on the wall shear stress.}

\begin{figure}[H]
 \begin{center}
      \includegraphics[width=6.1cm]{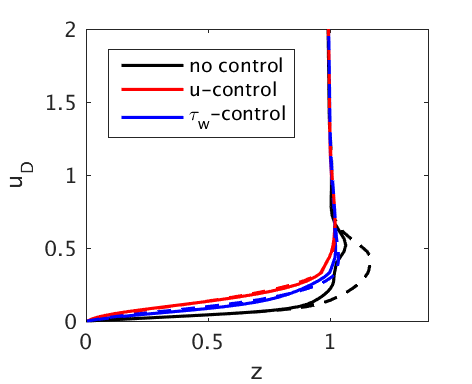}
      \includegraphics[width=6.1cm]{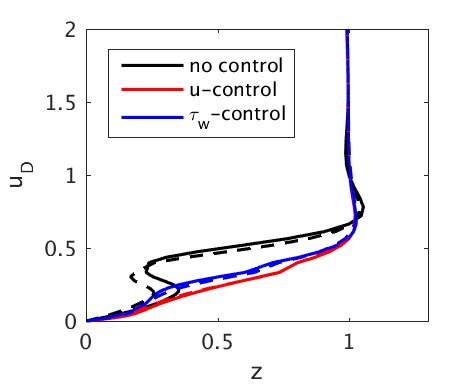}\\
a) \hspace{70mm} b) \\
      \includegraphics[width=6.1cm]{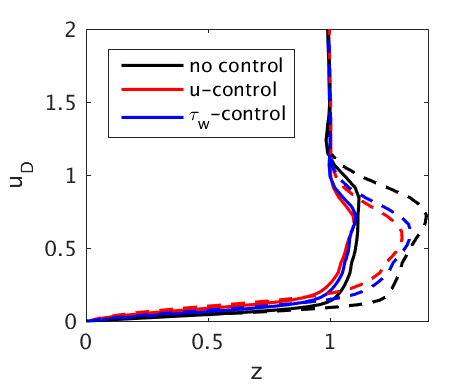}
      \includegraphics[width=6.1cm]{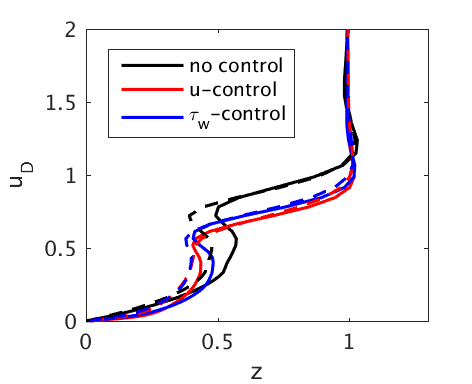}\\
c) \hspace{70mm} d) 
 \end{center}
  \caption{\label{} Profiles of streamwise velocity for isothermal (solid line) and adiabatic (dashed line) wall condition: a) $M_{\infty} = 1.5$, between two mushroom shapes; c) $M_{\infty} = 1.5$, in the center of a mushroom; d) $M_{\infty} = 7.5$, between two mushroom shapes; e) $M_{\infty} = 7.5$, in the center of a mushroom.}
  \label{f10}
\end{figure}

Typical streamwise velocity profiles taken between two mushroom shapes and in the center of a mushroom are shown in figure \ref{f10} for the lowest and the highest Mach numbers, $1.5$ and $7.5$, respectively, where the streamwise location coincides with the location {\color{black} where the} energy saturation {\color{black} begins}. For the adiabatic uncontrolled case ({\color{black} black} dashed line) a jet {\color{black} forms roughly at the vertical location $y=0.4$ for $M=1.5$ and $y=0.7$ for $M=7.5$}, in between two mushroom shapes, becoming more intense for the highest Mach number. The intensity of the jet {\color{black} is} weakened {\color{black} by both types of} the control (black and red curves in figures \ref{f8}b and \ref{f8}d). The right column shows streamwise velocity profiles at the center of the mushroom ($z=0$); for the lower Mach number there is no significant difference between the adiabatic and the isothermal wall conditions, but at higher Mach number the figure shows that the height of the mushroom shape is larger in adiabatic conditions. This is commensurate with the boundary layer thickness increase due to the {\color{black} high level of} heating effects. The effect of the control on the velocity profiles at this spanwise location $z=0$ is a flattening of the curves.

\subsection{Effect of wall heat transfer}

In this section, we scrutinize the effect of a passive control technique based on wall cooling {\color{black} and heating,} for the same Mach numbers {\color{black} and flow conditions} considered in the previous section. To this end, an isothermal boundary layer is considered and disturbed in the same manner as in the previous section, i.e., using an inflow disturbance imposed over a given base flow. The wall temperature at the inflow is set to the ambient temperature ($300$ K). The spanwise separation of the vortices - dictated by the spanwise wavenumber of the incoming disturbance and is set the same for all Mach numbers. Uniform wall cooling {\color{black} and heating}, corresponding to four temperatures $T_w = 150, 200, 400, 500$ K is applied starting at the streamwise location $x=8$. {\color{black} Note that the wall cooling or heating in this analysis refers to a decrease or an increase, respectively, of the upstream (base) wall temperature, which is equal to the ambient temperature, $300$ K. This is somewhat different to what was previously considered cooling or heating of the wall with respect to an adiabatic wall. For example, the effect of wall cooling with respect to the adiabatic condition can be noticed in figures \ref{f6} or \ref{f7}, by comparing the solid lines to the dashed lines, which shows that the wall cooling increases both the vortex energy and the wall shear stress, which is consistent with the previous studies, such as Spall and Malik \cite{Spall} or Elliot and Bassom \cite{Elliot}.} Results in terms of streamwise velocity contours, vortex energy distribution and spanwise averaged wall shear stress will now be presented and discussed.

\begin{figure}[H]
 \begin{center}
      \includegraphics[width=5.2cm]{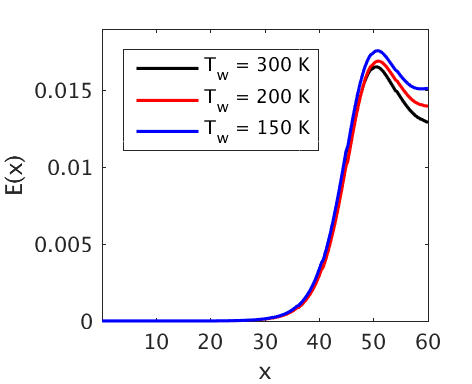}
      \includegraphics[width=5.2cm]{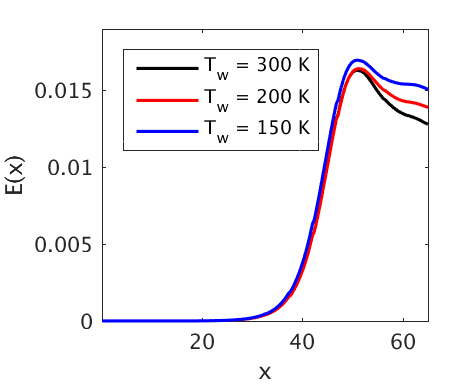}
      \includegraphics[width=5.2cm]{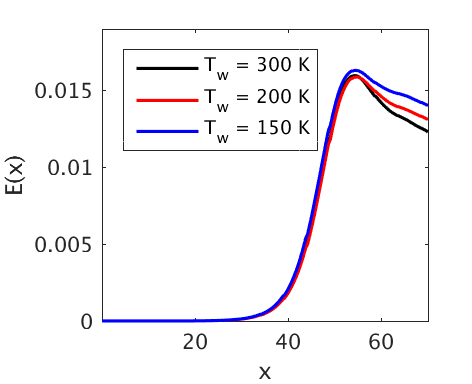}\\
a) \hspace{50mm} b) \hspace{50mm} c) \\
      \includegraphics[width=5.2cm]{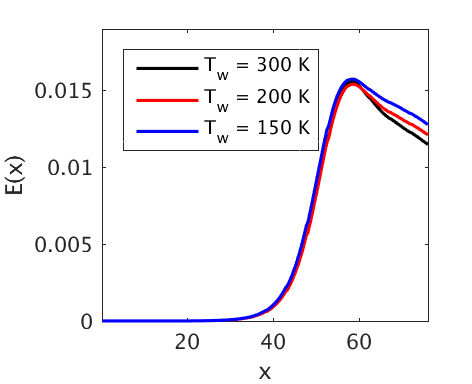}
      \includegraphics[width=5.2cm]{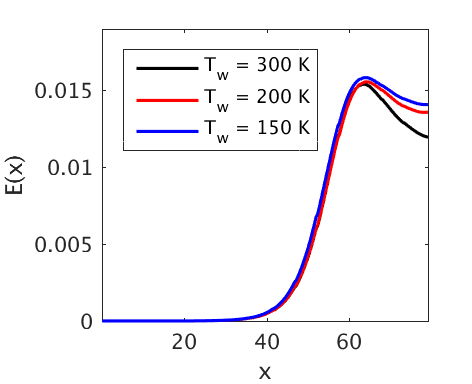}\\
d) \hspace{60mm} e) 
 \end{center}
  \caption{\label{} Energy of the disturbance as a function of the streamwise coordinate for wall cooling: a) $M_{\infty} = 1.5$ b) $M_{\infty} = 3.0$; c) $M_{\infty} = 4.5$; d) $M_{\infty} = 6.0$; e) $M_{\infty} = 7.5$.}
  \label{f11}
\end{figure}

\begin{figure}[H]
 \begin{center}
      \includegraphics[width=5.2cm]{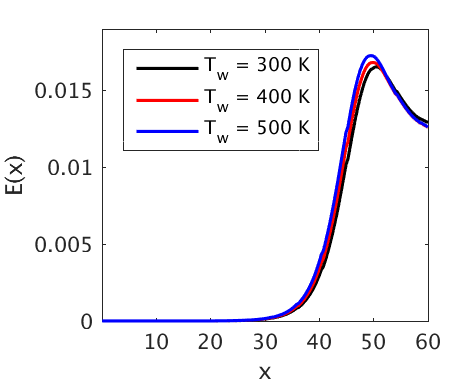}
      \includegraphics[width=5.2cm]{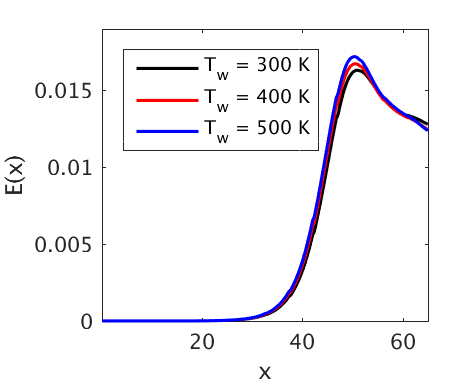}
      \includegraphics[width=5.2cm]{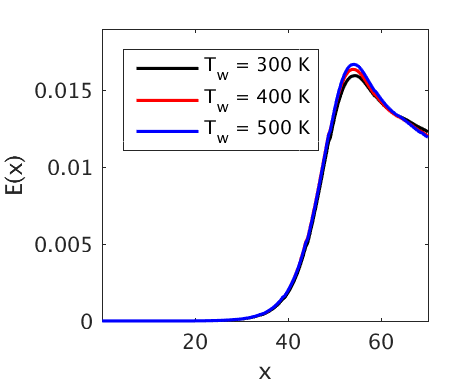}\\
a) \hspace{50mm} b) \hspace{50mm} c) \\
      \includegraphics[width=5.2cm]{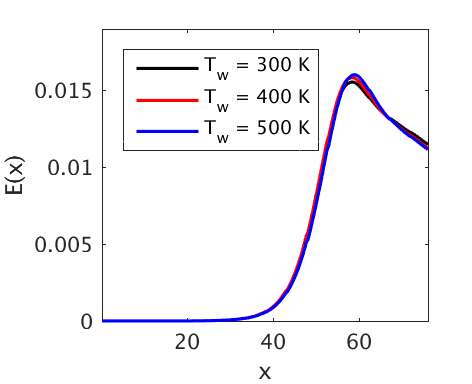}
      \includegraphics[width=5.2cm]{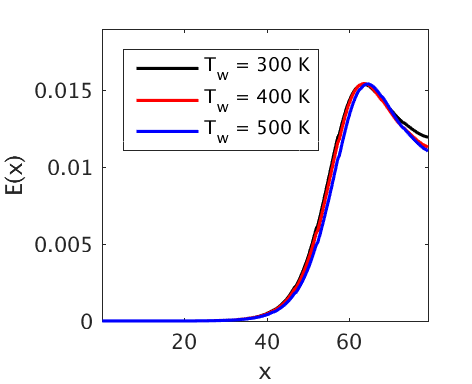}\\
d) \hspace{60mm} e) 
 \end{center}
  \caption{\label{} {\color{black} Energy of the disturbance as a function of the streamwise coordinate for wall heating: a) $M_{\infty} = 1.5$ b) $M_{\infty} = 3.0$; c) $M_{\infty} = 4.5$; d) $M_{\infty} = 6.0$; e) $M_{\infty} = 7.5$.}}
  \label{f12}
\end{figure}

The small increase of vortex energy by wall cooling or heating is revealed in figures \ref{f11} and \ref{f12}, {\color{black} respectively}, where distributions of {\color{black} vortex} energy with respect to the streamwise directions are {\color{black} superposed; note that these curves are not plotted in a logarithmic scale in order to reveal the small difference among the curves in the saturation region. While the energy saturation starting point is approximately the same for all cooling or heating cases, some differences among the energy levels can be noticed after the saturation point when wall cooling in applied. It appears that this level of wall heating (which is not very significant, anyway) has very little effect on the energy distribution as seen in figure \ref{f12}.}

\begin{figure}[H]
 \begin{center}
      \includegraphics[width=5.2cm]{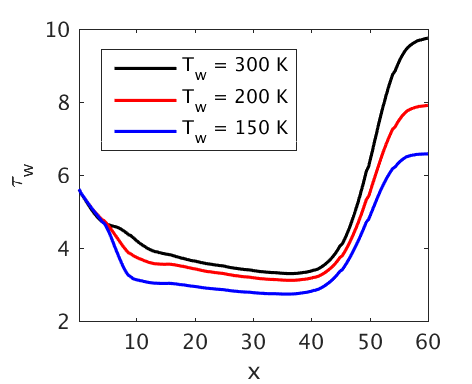}
      \includegraphics[width=5.2cm]{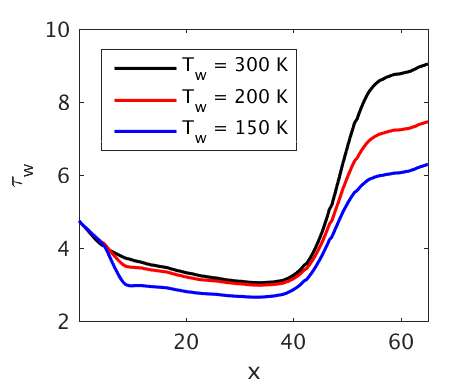}
      \includegraphics[width=5.2cm]{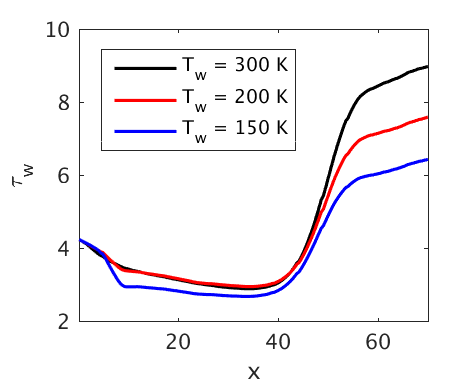}\\
a) \hspace{50mm} b) \hspace{50mm} c) \\
      \includegraphics[width=5.2cm]{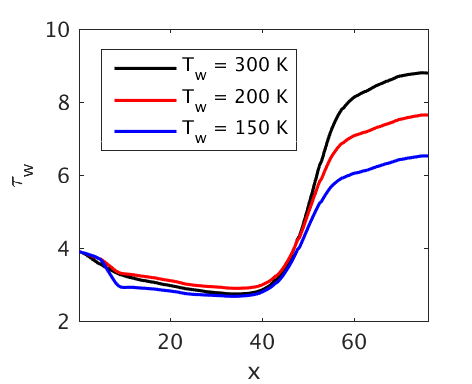}
      \includegraphics[width=5.2cm]{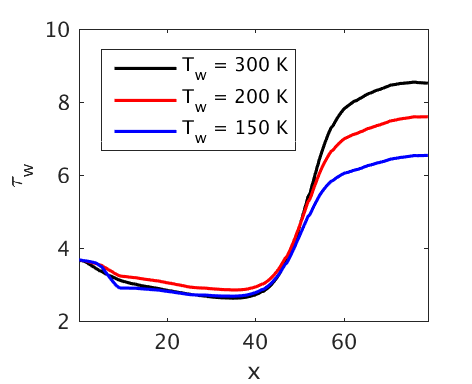}\\
d) \hspace{60mm} e) 
 \end{center}
  \caption{\label{} Spanwise average wall shear stress as a function of the streamwise coordinate for wall cooling: a) $M_{\infty} = 1.5$ b) $M_{\infty} = 3.0$; c) $M_{\infty} = 4.5$; d) $M_{\infty} = 6.0$; e) $M_{\infty} = 7.5$.}
  \label{f13}
\end{figure}

\begin{figure}[H]
 \begin{center}
      \includegraphics[width=5.2cm]{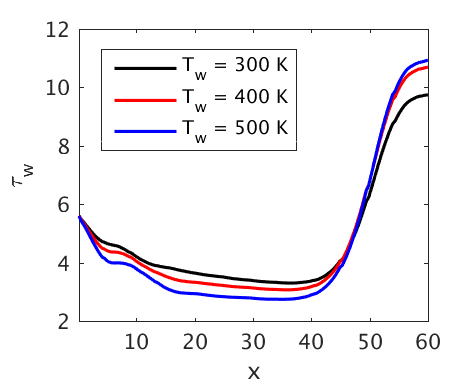}
      \includegraphics[width=5.2cm]{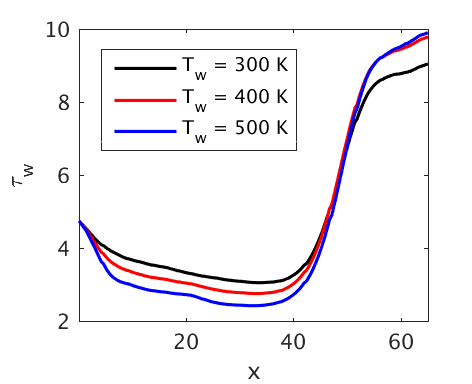}
      \includegraphics[width=5.2cm]{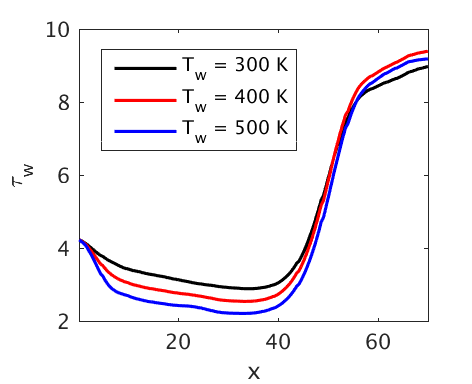}\\
a) \hspace{50mm} b) \hspace{50mm} c) \\
      \includegraphics[width=5.2cm]{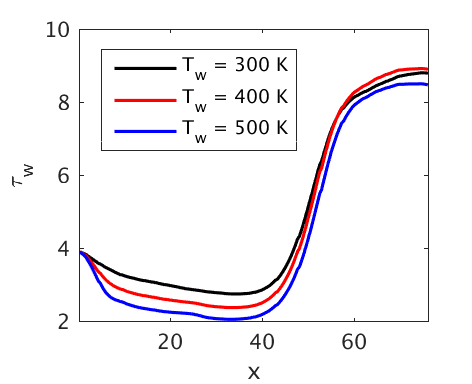}
      \includegraphics[width=5.2cm]{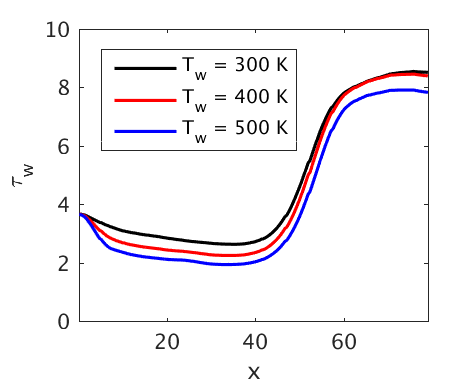}\\
d) \hspace{60mm} e) 
 \end{center}
  \caption{\label{} {\color{black} Spanwise average wall shear stress as a function of the streamwise coordinate for wall heating: a) $M_{\infty} = 1.5$ b) $M_{\infty} = 3.0$; c) $M_{\infty} = 4.5$; d) $M_{\infty} = 6.0$; e) $M_{\infty} = 7.5$.}}
  \label{f14}
\end{figure}

In figure \ref{f13} and \ref{f14}, the spanwise averaged wall shear stress is plotted as a function of the streamwise coordinate aiming to quantify the effect that the wall cooling {\color{black} and heating} may have on the skin friction. For all Mach number cases in figure \ref{f13} there is a reduction in the wall shear stress along the streamwise direction at the two lower wall temperatures, $150$ and $200$ K, {\color{black} compared to the original wall temperature of $300$ K}. {\color{black} As far as the wall heating is concerned, however, figure \ref{f14} suggests that for supersonic Mach numbers ($1.5$, $3$ and $4.5$) there is a slight shear stress reduction upstream of $x=50$ followed by a slight increase. For hypersonic Mach numbers ($6$ and $7.5$), however, there is a slight shear stress decrease everywhere. This rather curious result will be further investigated in a future study.} 

\begin{figure}[H]
 \begin{center}
      \includegraphics[width=7cm]{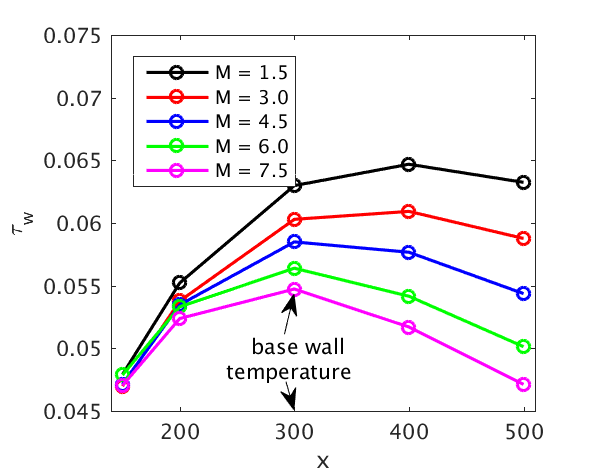} 
 \end{center}
  \caption{\label{} Skin friction drag as a function of the wall temperature and for different Mach numbers.}
  \label{f15}
\end{figure}

{\color{black} To quantify the overall frictional drag, we integrate the wall shear stress over the entire wall surface, and plot its variation with respect to the Mach number in figure \ref{f15}, where we mark the base wall temperature at $T=300$ K, which was imposed in the upstream of the cooling or heating region. The figure show that reducing the wall temperature (cooling) has a beneficial effect on the skin friction drag for all Mach numbers corresponding to both supersonic and hypersonic regimes. Increasing the wall temperature (heating) in the downstream of the base temperature still has a beneficial effect for high Mach number cases ($4.5$, $6$ and $7.5$), but insignificant effect on the other low Mach numbers ($1.5$ and $3$); there is actually a slight increase followed by a decrease of the drag for the lowest Mach number.}


\section{Conclusions}

A numerical study of the effect of controlled wall transpiration, cooling, and heating on the development of G\"{o}rtler vortices in high-speed boundary layers has been performed. For wall transpiration, a proportional controller was employed, where the sensors were either the streamwise velocity disturbance distribution in $x$ and $z$, {\color{black} at a specified distance from the wall,} or the wall shear stress distribution. {\color{black} While the first type of control is rather impractical since the sensors should be placed in the boundary layer flow, the second type is more amenable since the sensors are installed at the wall, so the interference with the flow is minimized.} Both isothermal and adiabatic wall conditions were considered for the control based on wall transpiration. The wall cooling {\color{black} or heating} was implemented in the framework of a passive control scheme, {\color{black} where the wall temperature was switched from the base temperature of $T=300$ imposed in the proximity to the inflow boundary to a lower or higher temperature in the downstream direction. The switching of wall temperature was performed using a ramping function to avoid discontinuities in the temperature at the wall. This is unlike previous studies where the wall cooling was considered with respect to an adiabatic wall condition. But in this case, even a wall temperature equal to the ambient temperature provides cooling because the adiabatic wall temperature is greater than the ambient temperature.} Both supersonic and hypersonic boundary layers, with Mach number ranging from $1.5$ to $7.5$, were investigated using a high-order numerical algorithm that solves the full Navier-Stokes equations in generalized curvilinear coordinates.

Results in terms of the density or streamwise velocity contours, vortex energy, spanwise averaged wall shear stress, and streamwise velocity profiles clearly indicated that both the wall transpiration and cooling/heating modify the streamwise {\color{black} evolution of the G\"{o}rtler vortices} for a given Mach number. {\color{black} It was found that the control based on sensors placed in the flow is more effective in reducing the vortex energy or the wall shear stress than the control based on sensors placed at the wall. The level of vortex energy reduction as seen in figure \ref{f6} seems to be roughly the same between the isothermal and adiabatic wall conditions.} The effect of wall transpiration on the reduction of G\"{o}rtler vortex energy is not as high as in the incompressible regime (see results reported in Sescu et al. \cite{Sescu1,Sescu2}). {\color{black} This may be explained by the reduced ability of a high speed flow to accommodate according to the inputs provided by sensors, and due to the compressibility effects.}

It was found that wall cooling {\color{black} slightly} increases the vortex energy {\color{black} in the saturation region} for all Mach numbers (figure \ref{f11}), although the spanwise averaged wall shear stress decreased (figure \ref{f13}). An energy increase may accelerate the onset of secondary instabilities and encourage early transition to turbulence. {\color{black} For all Mach numbers cases, the two lower wall temperatures, $150$ and $200$ K, reduce the wall shear stress along the streamwise direction compared to the base wall temperature of $300$ K that was imposed in the upstream. Figure \ref{f14} showed that wall heating for supersonic Mach numbers ($1.5$, $3$ and $4.5$) slightly reduces the wall shear stress upstream of $x=50$ followed by a slight increase. For hypersonic Mach numbers ($6$ and $7.5$), however, there was a slight decrease of the wall shear stress along the entire boundary layer. }

\section*{Acknowledgments}

 The authors would like to gratefully acknowledge constructive corrections, comments
and suggestions from anonymous reviewers.
A.S. would like to acknowledge partial support from NASA EPSCoR RID Program through Mississippi State Grant Consortium directed by Dr. Nathan Murray.
M.Z.A. would like to thank Strathclyde University for financial
support from the Chancellor's Fellowship.



\begin{thebibliography}{9}

 \bibitem{choudhari}
Choudhari, M. and Fischer, P. (2006) Roughness induced Transient Growth: Nonlinear Effects, In: Govindarajan R. (eds) IUTAM Symposium on Laminar-Turbulent Transition. Fluid Mechanics and Its Applications, vol 78. Springer, Dordrecht.

 \bibitem{white1}
White, E.B. (2002) Transient growth of stationary disturbances in a flat plate boundary layer, {\it Phys. Fluids}, Vol. 14, pp. 4429-4439.

 \bibitem{white2}
White, E.B., Rice, J.M. and Ergin, F.G. (2005) Receptivity of stationary transient disturbances to surface roughness, {\it Physics of Fluids}, Vol. 17, pp. 064109.
 
 \bibitem{Goldstein1}
Goldstein, M., Sescu, A., Duck, P. and Choudhari, M. (2010) The Long Range Persistence of Wakes behind a Row of Roughness Elements, {\it Journal of Fluid Mechanics}, Vol. 644, pp. 123-163.

 \bibitem{Goldstein2}
Goldstein, M., Sescu, A., Duck, P. and Choudhari, M. (2011) Algebraic/transcendental Disturbance Growth behind a Row of Roughness Elements, {\it Journal of Fluid Mechanics}, Vol. 668, pp. 236-266.

 \bibitem{Goldstein3}
Goldstein, M., Sescu, A., Duck, P. and Choudhari, M. (2016) Nonlinear wakes behind a row of elongated roughness elements, {\it Journal of Fluid Mechanics}, Vol. 796, pp. 516-557.

\bibitem{Ruban}
Ruban, A.I. \& Kravtsova, M.A. (2013) Generation of steady longitudinal vortices in hypersonic boundary layer. {\it J. Fluid Mech.}, Vol. 729, pp. 702-731.

 \bibitem{Kendall}
Kendall, J.M. (1998) Experiments on boundary-layer receptivity to freestream turbulence. {\it AIAA Paper 2004-2335}.

 \bibitem{Westin}
Westin, K.J.A., Boiko, A.V., Klingmann, B.J.B., Kozlov, V.V. \& Alfredsson, P.H. (1994) Experiments in a boundary layer subjected to free stream turbulence. part 1. Boundary layer structure and receptivity. {\it J. Fluid Mech.}, Vol. 281, pp. 193-218.

 \bibitem{wu1}
Wu, X. and Choudhari, M. (2011) Linear and nonlinear instabilities of a blasius boundary layer perturbed by streamwise vortices. Part 2. Intermittent instability induced by long wavelength Klebanoff modes. {\it J. Fluid Mech.}. Vol. 483, pp. 249-286.

 \bibitem{Matsubara}
Matsubara, M. and Alfredsson, P.H. (2001) Disturbance growth in boundary layers subjected to free stream turbulence, {\it J. Fluid Mech.}, Vol. 430, pp. 149.
 
 \bibitem{Leib}
Leib, S.J., Wundrow, W. \& Goldstein, M. (1999) Effect of free-stream turbulence and other vortical disturbances on a laminar boundary layer. {\it J. Fluid Mech.}, Vol. 380, pp. 169-203.

 \bibitem{Jacobs}
Jacobs, R.G. and Durbin, P. (2001) Simulations of bypass transition, {\it J. Fluid Mech.}, Vol. 428, pp. 185-212.

 \bibitem{Zaki}
Zaki, T.A. \& Durbin, P. (2005) Mode interaction and the bypass route to transition. {\it J. Fluid Mech.}, Vol. 531, pp. 85-111.

 \bibitem{Goldstein4}
Goldstein, M. \& Sescu (2008) Boundary-layer transition at high free-stream disturbance levels - beyond Klebanoff modes. {\it J. Fluid Mech.}, Vol. 613, pp. 95-124.

 \bibitem{ricco}
{\color{black} Ricco, P. Luo, J. Wu, X. (2011) Evolution and instability of unsteady nonlinear streaks generated by free-stream vortical disturbances, {\it J. Fluid Mech.}, Vol. 677, pp. 1-38.}

 \bibitem{Marensi}
{\color{black} Marensi E., Ricco P., Wu X. (2017) Nonlinear unsteady streaks engendered by the interaction of free-stream vorticity with a compressible boundary layer, {\it J. Fluid Mech.}, Vol: 817, pp. 80-121.}

 \bibitem{Landahl}
Landahl, M.T. (1980) A note on an algebraic instability of inviscid parallel shear flows. {\it J. Fluid Mech.}, Vol. 98, pp. 243-251.

 \bibitem{Gortler}
G\"{o}rtler, H . (1941) Instabilita-umt laminarer Grenzchichten an Konkaven Wanden gegenber gewissen dreidimensionalen Storungen, {\it ZAMM}, Vol. 21, pp. 250--52; english version: {\it NACA Report 1375} (1954).

 \bibitem{hall1}
Hall, P. (1982) Taylor-G\"{o}rtler vortices in fully developed or boundary-layer flows: linear theory, {\it Journal of Fluid Mechanics}, Vol. 124, pp. 475-494.

 \bibitem{hall2}
Hall, P. (1983) The linear development of G\"{o}rtler vortices in growing boundary layers, {\it Journal of Fluid Mechanics}, Vol. 130, pp. 41-58.

 \bibitem{hall3}
Hall, P. and Horseman. N. (1991) The linear inviscid secondary instability of longitudinal vortex structures in boundary layers, {\it Journal of Fluid Mechanics}. Vol. 232, pp. 357-375.

 \bibitem{Floryan}
Floryan, J.M. and Saric, W.S. (1982) Stability of G\"{o}rtler Vortices in Boundary Layers, {\it AIAA Journal}, Vol. 20, No. 3, pp. 316-324.

 \bibitem{Swearingen}
Swearingen, J.D. and Blackwelder, R.F. (1987) The growth and breakdown of streamwise vortices in the presence of a wall. {\it J. Fluid Mech.}, Vol. 182, pp. 255-290.

 \bibitem{Malik}
Malik, M.R. \& Hussaini, M.Y. (1990) Numerical simulation of interactions between G\"{o}rtler vortices and Tollmien-Schlichting waves. {\it J. Fluid Mech.}, Vol. 210, pp. 183-199.

 \bibitem{Saric}
Saric, W.S. (1994) G\"{o}rtler Vortices, {\it Annu. Rev. Fluid Mech.}, Vol. 26, pp. 379-409.

\bibitem{Li}
Li, F. and Malik, M. (1995) Fundamental and subharmonic secondary instabilities of G\"{o}rtler vortices, {\it Journal of Fluid Mechanics}, Vol. 297, pp. 77-100.

 \bibitem{Boiko}
Boiko, A.V., Ivanov, A.V., Kachanov, Y.S. and Mischenko, D.A. (2010) Steady and unsteady G\"{o}rtler boundary-layer instability on concave wall, {\it European Journal of Mechanics B/Fluids}, Vol. 29, pp. 61-83.

 \bibitem{wu}
Wu, X, Zhao, D. and Luo, J (2011) Excitation of steady and unsteady G\"{o}rtler vortices by free-stream vortical disturbances, {\it Journal of Fluid Mechanics}, Vol. 682, pp. 66-100.


 \bibitem{Sescu3}
Sescu, A. and Thompson, D. (2015) On the Excitation of G\"{o}rtler Vortices by Distributed Roughness Elements, {\it Theoretical and Computational Fluids Dynamics}, Vol. 29, pp. 67-92.

 \bibitem{Dempsey}
Depmsey, L.D., Hall, P. \& Deguchiu, K. (2017) The excitation of G\"{o}rtler vortices by free stream coherent structures, {\it J. Fluid Mech.}, Vol. 826, pp. 60-96.

 \bibitem{Xu}
Xu, D., Zhang, Y. \& Wu, X. (2017) Nonlinear evolution and secondary instability of steady and unsteady G\"{o}rtler vortices induced by free-stream vortical disturbances, {\it J. Fluid Mech.}, Vol. 829, pp. 681-730.

 \bibitem{El-Hady}
El-Hady, N.M. and Verma, A.K. (1984) Gortler Instability of Compressible Boundary Layers, {\it AIAA Journal}, Vol. 22, No. 10, pp. 1354-1355.

 \bibitem{Chen}
Chen, F.-J., Malik, M.R. and Beckwith, I.E. (1992) G\"{o}rtler instability and supersonic quiet nozzle design, {\it AIAA Journal}, Vol. 30, No. 8, pp. 2093-2094.

 \bibitem{hall4}
{\color{black} Hall, P. and Fu, Y. (1989) On the G\"{o}rtler vortex instability mechanism at hypersonic speeds {\it Theoretical and Computational Fluid Dynamics}. Vol. 1, pp. 125-134.}

 \bibitem{Fu}
{\color{black} Fu, Y. and Hall, P. (1993) Effects of G\"{o}rtler vortices, wall cooling and gas dissociation on the Rayleigh instability in a hypersonic boundary layer, {\it Journal of Fluid Mechanics}. Vol. 247, pp. 503-525.}

 \bibitem{Dando}
{\color{black} Dando, A.H. and Seddougui, S.O. (1993) The compressible G\"{o}rtler problem in two-dimensional boundary layers, {\it IMA J. Appl. Math.}. Vol. 51, pp. 27-67.}

 \bibitem{Ren}
Ren, J. \& Fu, S. (2017) Secondary instabilities of G\"{o}rtler vortices in high-speed boundary layer flows, {\it J. Fluid Mech.}, Vol. 781, pp. 388-421.

 \bibitem{Choi}
Choi, H., Moin, P. and Kim, J. (1994) Active turbulence control for drag reduction in wall-bounded flows. {\it J. Fluid Mech.}, Vol. 262, pp. 75-110.

 \bibitem{Koumoutsakos1}
Koumoutsakos, P. (1997) Active control of vortex-wall interactions, {\it Phys. Fluids}, Vol. 9, pp. 3808. 

 \bibitem{Koumoutsakos2}
Koumoutsakos, P. (1999) Vorticity flux control for a turbulent channel flow, {\it Phys. Fluids}, Vol. 11, pp. 248.
 
  \bibitem{Lee}
 Lee, C., Kim, J. \& Choi, H. (1998) Suboptimal control of turbulent channel flow for drag reduction. {\it J. Fluid Mech.}, Vol. 358, pp. 245-258.

 \bibitem{Pamies}
Pamies, M., Garnier, E., Merlen, A. \& Sagaut, P. (2007) Response of a spatially developing turbulent boundary layer to active control strategies in the framework of opposition control. {\it Phys. Fluids}, Vol. 19, pp. 108102.

 \bibitem{Stroh}
Stroh, A., Frohnapfel, B., Schlatter, P. \& Hasegawa, Y. (2015) A comparison of opposition control in turbulent boundary layer and turbulent channel flow. {\it Phys. Fluids}, Vol. 27, pp. 075101.

 \bibitem{Kim}
Kim, J. (2003) Control of turbulent boundary layers. {\it Phys. Fluids}, Vol. 15, pp. 1093.

 \bibitem{Floryan2}
Floryan, J.M. and Saric, W.S. (1983) Effects of suction on the G\"{o}rtler instability of boundary layers, {\it AIAA Journal}, Vol. 21, No. 12, pp. 1635-1639.

 \bibitem{Floryan3}
Floryan, J.M. (1985) Effects of blowing on the G\"{o}rtler instability of boundary layers, {\it AIAA Journal}, Vol. 23, No. 8, pp. 1287-1288.

 \bibitem{El-Hady2}
El-Hady, N.M. and Verma, A.K. (1984) Instability of compressible boundary layers along curved walls with suction or cooling, {\it AIAA Journal}, Vol. 22, No. 2, pp. 206-213.

 \bibitem{Xu2}
 Xu, C.-X., Choi, J. and Sung, H.J. (2003) Identification and Control of Taylor-Gortler Vortices in Turbulent Curved Channel Flow, {\it AIAA Journal}, Vol. 41, No. 12, pp. 2387-2393.

 \bibitem{Nishihara}
Nishihara, M., Jiang, N., Rich, J.W., Lempert, W.R. \& Adamovich, V. (2005) Low-temperature supersonic boundary layer control using repetitively pulsed magnetohydrodynamic forcing, {\it Phys. Fluids}, Vol. 17, pp. 106102.

 \bibitem{Im}
Im, S., Do, H. \& Cappelli, M.A. (2010) Dielectric barrier discharge control of a turbulent boundary layer in a supersonic flow, {\it Phys. Fluids}, Vol. 97, pp. 041503.

 \bibitem{Xu2}
Xu, G., Xiao, Z. \& Fu, S. (2011) Secondary instability control of compressible flow by suction for a swept wing, Physics, Mechanics \& Astronomy, Vol. 54, pp. 2040-2052.

 \bibitem{Sun}
Sun, Z., Ren, Y. \& Larricq, L. (2011) Drag reduction of compressible wall turbulence with active dimples, Physics, Mechanics \& Astronomy, Vol. 54, pp. 239-337.

 \bibitem{Blinde}
Blinde, P.L., Humble,R.A., Oudheusden,B.W., Scarano, F. (2009) Effects of micro-ramps on a shockwave/turbulent boundary layer interaction. Shock Waves 19, 507-520.


\bibitem{Pasquariello}
Pasquariello,V., Grilli, M.,Hickel, S., Adams, N. (2014) Large-eddy simulation of passive shock-wave/boundary-layer interaction control. Int. J. Heat Fluid Flow 49, 116-127.

 \bibitem{Fukuda}
Fukuda, M.K., Roshotko, E., Higst, W.R. (1977) Bleed effects on shock/boundary-layer interactions in supersonic mixed compression inlets. J. Aircraft 14(2), 151-156.

 \bibitem{Vadillo}
Vadillo, J.L., Agarwal, R.K., Hassan, A.A. (2006) Active Control of Shock/Boundary Layer Interaction in Transonic Over Airfoils. In: Groth C., Zingg D.W. (eds) Computational Fluid Dynamics 2004. Springer, Berlin, Heidelberg.

 \bibitem{Verma}
Verma, S.B. \& Hadjadj, A. (2015) Supersonic flow control, Shock Waves, Vol. 25, pp. 443-449.

 \bibitem{Fedorov}
Fedorov, A.V. (2015) Prediction and control of laminar-turbulent transition in high-speed boundary layers, Procedia IUTAM, Vol. 14, pp. 3-14.

 \bibitem{Joshi}
Joshi, S. S., Speyer, L., J. and Kim, J. 1997 A systems theory approach to the feedback stabilization of infinitesimal and finite-amplitude disturbances in plane Poiseuille flow. {\it J. Fluid Mech.}, Vol. 332, pp. 157-184.

 \bibitem{Hanson}
Hanson, R. E., Bade, K. M., Belson, B. A., Lavoie, P., Naguib, A. M. and Rowley, C. W. (2014) Feedback control of slowly-varying transient growth by an array of plasma actuators. {\it Phys. Fluids}, Vol. 26, pp. 024102.

 \bibitem{Jacobson}
Jacobson, S. A. and Reynolds, W. C. (1998) Active control of streamwise votices and streaks in boundary layers. {\it J. Fluid Mech.}, Vol. 360, pp. 179-211.

 \bibitem{Tannehill}
Tannehill, J.C, Anderson, D.A. and Pletcher, R.H. (1997) Computational Fluid Mechanics and Heat Transfer, 2nd Ed., Taylor \& Francis, 1997, ISBN 9781591690375.

\bibitem{Liu}
 Liu. X., Osher, S. and Chan, T. (1994) {Weighted essentially non-oscillatory schemes}, Journal of Computational Physics, Vol. 115, pp. 200-212.

\bibitem{Tam}  
Tam, C.K.W. and Webb, J.C. (1993), Dispersion-relation-preserving finite difference schemes for Computational Aeroacoustics, {\it Journal of Computational Physics}, Vol. 107, pp. 262-281.

 \bibitem{Schrader}
Schrader, L., Brandt, L. and Zaki, T.A. (2011) Receptivity, instability and breakdown of Gortler flow, {\it J. Fluid Mech.}, Vol. 682, pp. 362-396.

\bibitem{Xu2}  
Xu, D., Zhang, Y. and Wu, X. (2017) Nonlinear evolution and secondary instability of steady and unsteady G\"{o}rtler vortices induced by free-stream vortical disturbances, {\it J. Fluid Mech.}, Vol. 829, pp. 681-730.

 \bibitem{Ziegler}
Ziegler, M. and Nichols, N.B. (1942) Optimal settings for automatic controllers, {\it Trans. ASME}, Vol. 65, pp. 433-444.

\bibitem{Li2}
Li, F., Choudhari, M., Chang, C.-L., Greene, P., and Wu, M. (2010) Development and Breakdown of Gortler Vortices in High Speed Boundary Layers, 48th AIAA Aerospace Sciences Meeting Including the New Horizons Forum and Aerospace Exposition, Aerospace Sciences Meetings.

 \bibitem{Sescu1}
Sescu, A., Taoudi, L. and Afsar, M. (2018) Iterative control of G\"{o}rtler vortices via local wall deformations, {\it Theoretical and Computational Fluids Dynamics}, Vol. 32, pp. 63-72.

 \bibitem{Paredes}
Paredes, P., Choudhari, M.M. and Li, F. (2016) Transition due to streamwise streaks in a supersonic flat plate boundary layer, {\it Physical Review Fluids}, Vol. 1, pp. 083601.

 \bibitem{Sescu2}
{\color{black} Sescu, A. and Afsar, M. (2018) Hampering G\"{o}rtler vortices via optimal control in the framework of nonlinear boundary region equations, {\it J. Fluid Mech.}, Vol. 848, pp. 5-41.}

 \bibitem{Spall}
{\color{black} Spall, R.E. and Malik, M.R. (1989) G\"{o}rtler vortices in supersonic and hypersonic boundary layers, {\it Physics of Fluids A: Fluid Dynamics}, Vol. 1, pp. 1822.

 \bibitem{Elliot}
Elliot, J.W. and Bassom, A.P. (2000) The effect of wall cooling on compressible G\"{o}rtler vortices, {\it Eur. J. Mech. B - Fluids}, Vol. 19, pp. 37-68.}


\end{thebibliography}
\end{document}